# Fitting Large Nonlinear Mixed Effects Models Using Variational Expectation Maximization

**Mohamed Tarek**
PumasAI, USA
mohamed@pumas.ai

**Pedro Afonso**
PumasAI, USA
pedro.a@pumas.ai

## Abstract

Nonlinear Mixed Effects (NLME) models are widely used in pharmacometrics and related fields to analyze hierarchical and longitudinal data. However, as the number of parameters and random effects increases, traditional methods for maximizing the marginal likelihood become computationally expensive. This paper explores the *Variational Expectation Maximization (VEM)* algorithm, a scalable alternative for fitting NLME models. Originally introduced in the context of probabilistic graphical models and later popularized through variational autoencoders, VEM has not been extensively applied to NLME modeling. By leveraging flexible variational families and reverse-mode automatic differentiation, VEM can efficiently maximize the marginal likelihood, scaling to NLME models with over 15,000 population parameters. This work provides a detailed description of VEM, compares it to other NLME fitting algorithms, and highlights its scalability through computational experiments. Using the Pumas statistical software, we fit two test models: 1) a standard warfarin model, and 2) an unnecessarily over-parameterized DeepNLME Friberg model with 15,410 population parameters and 16 random effects. The warfarin model was fitted to completion to demonstrate the correctness of VEM, while the DeepNLME Friberg model instead demonstrates VEM's scalability on a toy but large model. VEM improves the log likelihood steadily over hundreds of iterations at a practical per-iteration cost, while FOCE fails to complete even one iteration within a day. The model is deliberately over-parameterized for its small dataset and over-fits it, so what this experiment establishes is that VEM optimizes the objective of a model of this size at a practical cost. Applying VEM to large models that are genuinely useful is left to future work.

## 1 Introduction

Nonlinear Mixed Effects (NLME) models are widely used in pharmacometrics and other fields to model data with hierarchical structure, such as longitudinal data from clinical trials. These models combine fixed effects (population-level parameters) and random effects (individual-level parameters), allowing us to reflect the hierarchical structure of the underlying data-generating process. Traditionally, the parameters of NLME models are estimated by maximizing the marginal likelihood. The marginal likelihood is the probability density (or mass) of the observed data given the population parameters, with the random effects integrated out. This is mathematically expressed as:

$$p(y \mid \theta) = \int p(y \mid \eta, \theta) \cdot p(\eta \mid \theta) \; d\eta = \int p(y, \eta \mid \theta) \; d\eta$$

where $y$ is the observed data, $\theta$ is the set of population parameters, and $\eta$ is the set of random effects. In this notation, $\theta$ includes the fixed effects but also any other parameters shared between all subjects, e.g. parameters of the distribution of the random effects or the residual error model.

In pharmacometrics, the conditional likelihood $p(y \mid \eta, \theta)$ typically requires solving a system of Ordinary Differential Equations (ODEs) that describe the drug's Pharmacokinetics (PK) and Pharmacodynamics (PD). In Quantitative Systems Pharmacology (QSP), the ODEs can be high dimensional, describing complex biological mechanisms with many parameters and interacting states. In DeepNLME models (Korsbo et al., 2023), $p(y \mid \eta, \theta)$ includes a neural network or a neural ODE, which can be expensive to evaluate and differentiate through. All of these model types are similar from a statistical perspective, as they all involve a set of parameters that are shared across subjects and a set of random effects that are specific to each subject. Evaluating the marginal likelihood requires integrating over the random effects in the model. Fitting the models to the data involves estimating the population parameters by maximizing the marginal likelihood with respect to the population parameters. This is a challenging problem because the marginal likelihood is generally not available in closed form for nonlinear models, and numerical methods are required to approximate it.

In the NLME literature, a number of methods have been developed to estimate the parameters of an NLME model by maximizing the marginal likelihood. Some methods scale better than others. However, for large models, all of the common NLME model fitting algorithms become computationally prohibitive. The likelihood-approximation methods First Order (FO), First Order Conditional Estimation (FOCE), and the Laplace method optimize an approximate marginal likelihood using its gradient with respect to the population parameters. When that gradient is computed with forward-mode Automatic Differentiation (AD) or finite differences, as is typical, its cost grows linearly with the number of population parameters, so all three scale poorly at large parameter counts. FO scales similarly to FOCE; although it is generally much faster because it skips the inner optimization over the random effects, it linearizes the model about the mean of the random effects (Sheiner & Beal, 1980) rather than about their conditional modes, so its approximation error grows with the model's nonlinearity in the random effects, and it may yield poor estimates as a result (Lindstrom & Bates, 1990), particularly when the residual or inter-individual variability is large (Bauer, 2019). The expectation–maximization methods Monte Carlo Expectation Maximization (MCEM) and Stochastic Approximation Expectation Maximization (SAEM) instead sample each subject's random effects by Markov Chain Monte Carlo (MCMC), which scales poorly in high dimensions when based on the random-walk Metropolis-Hastings algorithm (Hastings, 1970; Metropolis et al., 1953), whose efficiency degrades as the random-effect dimension grows (Beskos et al., 2013; Roberts et al., 1997). In this paper, we describe the *Variational Expectation Maximization (VEM)* algorithm, a scalable method for maximizing the marginal likelihood that can efficiently fit NLME models with more than 15,000 population parameters. The VEM algorithm was first proposed in the context of Probabilistic Graphical Models (PGMs) (Jordan et al., 1999). Later, a variant of the VEM algorithm was also used to train the well-known Variational Autoencoder (VAE) family of generative machine learning models. The VAE was independently introduced in two seminal papers: Kingma & Welling (2013) and Rezende et al. (2014). In the NLME context, variants of the VEM algorithm have also been used recently, in both amortized (Arruda et al., 2024; Baaz et al., 2026; Li et al., 2026) and non-amortized, per-subject (Janssen et al., 2024) forms. In this paper, we present a different variant of the VEM algorithm for NLME models; Section 4 positions it relative to these and other recent variational approaches in pharmacometrics.

The emphasis on large models in this paper is deliberate. For small models, FOCE and the expectation–maximization methods MCEM and SAEM already work quite well, and VEM is unlikely to be consistently faster than them in that regime. The value of VEM emerges when the model is large enough that these existing algorithms become too slow. This does not mean that VEM cannot be used for small models, but its added value there is likely to be limited or none.

The variant of VEM presented here is the one implemented in Pumas (Mogensen et al., 2026), a software platform for pharmacometric modeling and simulation written in the Julia programming language (Bezanson et al., 2017). Pumas provides a unified environment for specifying NLME models and estimating their parameters, supporting a range of estimation methods including FO, FOCE, the Laplace method, SAEM, full Bayesian estimation, and the new VEM algorithm described in this paper. Throughout, we describe VEM as implemented in Pumas, and the computational experiments in Section 5 use this implementation.

The rest of the paper is organized as follows. In Section 2, we discuss the expectation maximization family of algorithms. In Section 3, we present the Pumas variant of the VEM algorithm for NLME model fitting, including its connection to VAEs, Probabilistic Principal Component Analysis (PPCA), and the Laplace method (Bleistein & Handelsman, 1986). In Section 4, we position VEM relative to recent variational approaches in pharmacometrics. In Section 5, we present limited computational experiments to demonstrate that: 1) VEM works, and 2) it scales to large models. The computational experiments are not intended to be comprehensive benchmarks of VEM against competing methods and their software implementations. Instead, the primary focus of this paper is on the theoretical aspects of the VEM algorithm. In Section 6, we discuss the results and their implications. Finally, concluding remarks are provided in Section 7.

## 2 Expectation Maximization (EM)

In this section, the ideas behind the general EM algorithm (Dempster et al., 1977), MCEM algorithm (Wei & Tanner, 1990), and SAEM algorithm (Delyon et al., 1999; Kuhn & Lavielle, 2004; 2005) are reviewed in the context of NLME models. This is required background to understand the VEM algorithm presented in the next section.

### 2.1 Importance Sampling (IS)

Evaluating the marginal likelihood in closed form is generally not possible for nonlinear models. Therefore, numerical methods are required to approximate the integral. For high-dimensional $\eta$, the most straightforward way of doing so is by turning it into an expectation and then computing the average of some number (here $M$) of samples. That is:

$$
\begin{aligned}
p(y \mid \theta) &= \int p(y \mid \eta, \theta) \cdot p(\eta \mid \theta)\ d\eta \\
&= E_{p(\eta \mid \theta)} p(y \mid \eta, \theta) \\
&\approx \frac{1}{M} \sum_{j=1}^{M} p\big(y \mid \eta = \eta_j, \theta\big)
\end{aligned}
$$

where $\eta_j$ are samples drawn from the prior distribution $p(\eta \mid \theta)$. The issue with this approach is that it is inefficient. In general, $\eta$ samples from the prior may give low probability density to the observations, and the regions that contribute most to the marginal density may only be sampled occasionally. Ideally, we would like the sampling to focus on the regions with high density.

We can achieve this by multiplying and dividing by the probability density function (PDF) $q(\eta)$ of another distribution (to be determined), known as the proposal distribution, inside the integral. We can then formulate the expectation to use the new distribution (whose PDF is $q(\eta)$) as the sampling distribution in the expectation.

$$
p(y \mid \theta) = \int p(y, \eta \mid \theta)\ d\eta = \int \frac{p(y, \eta \mid \theta)}{q(\eta)} \cdot q(\eta)\ d\eta = E_{q(\eta)} \frac{p(y, \eta \mid \theta)}{q(\eta)}
$$

The associated estimator from a finite number of samples, $\hat{p}(y \mid \theta)$, can then be written as a weighted average (with the $j^{\text{th}}$ sample having weight/importance $w_j$) of the conditional likelihood $p\big(y \mid \eta = \eta_j, \theta\big)$, which is why this method is called IS (Kahn & Harris, 1951):

$$
\begin{aligned}
p(y \mid \theta) \approx \hat{p}(y \mid \theta) &= \frac{1}{M} \sum_{j=1}^{M} \frac{p\big(y, \eta = \eta_j \mid \theta\big)}{q\big(\eta = \eta_j\big)} \\
&= \frac{1}{M} \sum_{j=1}^{M} \underbrace{\frac{p\big(\eta = \eta_j \mid \theta\big)}{q\big(\eta = \eta_j\big)}}_{w_j} \cdot p\big(y \mid \eta = \eta_j, \theta\big) \\
&= \frac{1}{M} \sum_{j=1}^{M} w_j \cdot p\big(y \mid \eta = \eta_j, \theta\big)
\end{aligned}
$$

Here, the samples $\eta_j$ are obtained from the proposal distribution $q(\eta)$, and the weights $w_j$ are given by the ratio of the prior to the proposal evaluated at each sample, representing a correction factor that accounts for the fact that we are sampling from a different distribution than the prior. This is an unbiased estimator of the marginal likelihood, since:

$$
E_{\eta_1, \eta_2, \ldots, \eta_M \sim q(\cdot)} \left[ \frac{1}{M} \sum_{j=1}^{M} w_j \cdot p\big(y \mid \eta = \eta_j, \theta\big) \right] = E_{\eta_1, \eta_2, \ldots, \eta_M \sim q(\cdot)} \left[ \frac{1}{M} \sum_{j=1}^{M} \frac{p\big(y, \eta = \eta_j \mid \theta\big)}{q\big(\eta = \eta_j\big)} \right]
$$

$$
\begin{aligned}
&= E_{\eta \sim q(\cdot)} \frac{p(y, \eta \mid \theta)}{q(\eta)} \\
&= \int \frac{p(y, \eta \mid \theta)}{q(\eta)} q(\eta) \ d\eta \\
&= \int p(y, \eta \mid \theta) \ d\eta \\
&= p(y \mid \theta)
\end{aligned}
$$

However, for each choice of $q(\eta)$, the estimator has a different variance, and thus a different precision. The precision of the estimator is important because it determines how many samples $M$ are required to get a good estimate of the marginal likelihood. The optimal proposal distribution is one that minimizes the variance of the estimator, which is given by:

$$
\begin{aligned}
\mathrm{Var}_{\eta_1, \eta_2, \ldots, \eta_M \sim q(\cdot)}(\hat{p}(y \mid \theta)) &= \mathrm{Var}_{\eta_1, \eta_2, \ldots, \eta_M \sim q(\cdot)} \left( \frac{1}{M} \sum_{j=1}^{M} \frac{p(y, \eta = \eta_j \mid \theta)}{q(\eta = \eta_j)} \right) \\
&= \frac{1}{M} \cdot \mathrm{Var}_{\eta \sim q(\cdot)} \frac{p(y, \eta \mid \theta)}{q(\eta)}
\end{aligned}
$$

Therefore, to get a precise estimator (small estimator variance), either the variance of $p(y, \eta \mid \theta)/q(\eta)$ must be small or the sample size $M$ must be large. So the performance of IS for a given sample size depends on the ability of the proposal distribution to reduce the variance of $p(y, \eta \mid \theta)/q(\eta)$. From Bayes' rule:

$$
\begin{aligned}
p(\eta \mid y, \theta) &= \frac{p(y, \eta \mid \theta)}{p(y \mid \theta)} \\
\frac{p(y, \eta \mid \theta)}{p(\eta \mid y, \theta)} &= p(y \mid \theta) \to \text{constant in } \eta,
\end{aligned}
$$

the ratio $p(y, \eta \mid \theta)/q(\eta)$ is constant in $\eta$ (has zero variance with respect to $\eta$) when $q(\eta)$ is exactly equal to the posterior distribution $p(\eta \mid y, \theta)$. Therefore, the optimal proposal distribution $q(\eta)$ is the posterior distribution $p(\eta \mid y, \theta)$. The challenge with IS is that the posterior density is generally not known in closed form. However, an approximation can be obtained using Variational Inference (VI) (Blei et al., 2017; Jordan et al., 1999) or by fitting a parametric distribution to the posterior samples obtained from MCMC (Hastings, 1970; Metropolis et al., 1953; Robert & Casella, 2004) sampling.

### 2.2 Evidence Lower Bound (ELBO) for NLME models

In this section, we derive the ELBO for NLME models. The ELBO is a fundamental concept in EMs algorithms that enables the indirect maximization of the marginal likelihood by maximizing a lower bound on it. The ELBO is also the objective function that is maximized in VI and the VEM algorithm.

#### 2.2.1 NLME model structure

To make the model and data assumptions explicit, consider a study with $N$ subjects. For each subject $i$, let:

- $\theta$ be the set of population parameters,
- $\eta_i$ be the vector of subject-specific random effects, and
- $y_i$ be the vector of observed responses for subject $i$.

In longitudinal studies such as pharmacometric analyses, $y_i$ typically contains $n_i$ repeated measurements of one or more response variables (e.g., drug concentrations or biomarker levels) collected over time. The statistical model is built from two conditional distributions. First, the prior distribution of the random effects:

$$p(\eta_i \mid \theta)$$

which encodes between-subject variability. Second, the conditional likelihood of the observations:

$$p(y_i \mid \eta_i, \theta)$$

which encodes within-subject variability. In practice, this likelihood is constructed from a structural model and an observation model. The structural model gives the predicted mean/median response as a function of time, dose, covariates, and the subject-specific parameters; the observation model then specifies the distribution of the measurements around these predictions.

For each subject $i$, both the prior $p(\eta_i \mid \theta)$ and the conditional likelihood $p(y_i \mid \eta_i, \theta)$ are explicitly available as conditional distributions that we can sample from.

Collecting the random effects and observations of all $N$ subjects into $\eta$ and $y$, two assumptions structure the joint model: the random effects of different subjects are independent given the population parameters, and each subject's observations depend only on that subject's random effects and on the population parameters. That is,

$$p(\eta \mid \theta) = \prod_{i=1}^{N} p(\eta_i \mid \theta), \qquad p(y \mid \eta, \theta) = \prod_{i=1}^{N} p(y_i \mid \eta_i, \theta).$$

The joint probability therefore factorizes over subjects,

$$p(y, \eta \mid \theta) = p(\eta \mid \theta) \cdot p(y \mid \eta, \theta) = \prod_{i=1}^{N} p(y_i, \eta_i \mid \theta),$$

and marginalizing $\eta$ out factorizes likewise, so the population log marginal likelihood is a sum over subjects:

$$\log p(y \mid \theta) = \sum_{i=1}^{N} \log p(y_i \mid \theta).$$

### 2.2.2 ELBO derivation

Now, assume we are trying to maximize the above expression. To do so, we start by rewriting each individual log marginal likelihood as an expectation, as before:

$$
\begin{aligned}
\log p(y_i \mid \theta) &= \log \int p(y_i \mid \eta_i, \theta) \cdot p(\eta_i \mid \theta)\ d\eta_i \\
&= \log E_{p(\eta_i \mid \theta)} p(y_i \mid \eta_i, \theta)
\end{aligned}
$$

We can compute a lower bound by introducing an arbitrary proposal distribution $q_i(\eta_i)$ on the support of $p(\eta_i \mid \theta)$ and applying Jensen's inequality, which for a concave function $f$ (such as the logarithm) and a random variable $X$ says that:

$$
f(E[X]) \geq E[f(X)]
$$

In our case, this gives:

$$
\begin{aligned}
\log p(y_i \mid \theta) &= \log \int p(y_i \mid \eta_i, \theta) \cdot p(\eta_i \mid \theta)\ d\eta_i \\
&= \log \int p(y_i \mid \eta_i, \theta) \cdot \frac{p(\eta_i \mid \theta)}{q_i(\eta_i)} \cdot q_i(\eta_i)\ d\eta_i \\
&= \log \int \frac{p(y_i, \eta_i \mid \theta)}{q_i(\eta_i)} \cdot q_i(\eta_i)\ d\eta_i \\
&= \log E_{q_i(\eta_i)} \left[ \frac{p(y_i, \eta_i \mid \theta)}{q_i(\eta_i)} \right] \\
&\geq E_{q_i(\eta_i)} \left[ \log\left( \frac{p(y_i, \eta_i \mid \theta)}{q_i(\eta_i)} \right) \right] \\
&= \text{ELBO}_i
\end{aligned}
$$

The population log marginal likelihood can then be lower bounded by summing the individual lower bounds:

$$
\begin{aligned}
\text{ELBO} &= \sum_{i=1}^{N} \text{ELBO}_i \\
&\leq \sum_{i=1}^{N} \log p(y_i \mid \theta) = \log p(y \mid \theta)
\end{aligned}
$$

Using this identity:

$$
p(y_i, \eta_i, \theta) = p(\eta_i \mid y_i, \theta) \cdot p(y_i \mid \theta),
$$

we can rewrite the individual ELBO as:

$$
\begin{aligned}
\text{ELBO}_i &= E_{q_i(\eta_i)} \left[ \log\left( \frac{p(y_i, \eta_i \mid \theta)}{q_i(\eta_i)} \right) \right] \\
&= E_{q_i(\eta_i)} \left[ \log\left( \frac{p(\eta_i \mid y_i, \theta) \cdot p(y_i \mid \theta)}{q_i(\eta_i)} \right) \right]
\end{aligned}
$$

$$
\begin{aligned}
&= \log p(y_i \mid \theta) + E_{q_i(\eta_i)}\left[\log \frac{p(\eta_i \mid y_i, \theta)}{q_i(\eta_i)}\right] \\
&= \log p(y_i \mid \theta) - E_{q_i(\eta_i)}\left[\log \frac{q_i(\eta_i)}{p(\eta_i \mid y_i, \theta)}\right] \\
&= \log p(y_i \mid \theta) - \mathrm{KL}(q_i(\eta_i) \parallel p(\eta_i \mid y_i, \theta)) \qquad (1)
\end{aligned}
$$

where $\mathrm{KL}(q_i(\eta_i) \parallel p(\eta_i \mid y_i, \theta))$ is the Kullback-Leibler (KL) divergence between the proposal distribution $q_i(\eta_i)$ and the posterior distribution $p(\eta_i \mid y_i, \theta)$, which is defined as:

$$
\mathrm{KL}(q_i(\eta_i) \parallel p(\eta_i \mid y_i, \theta)) = E_{q_i(\eta_i)}\left[\log \frac{q_i(\eta_i)}{p(\eta_i \mid y_i, \theta)}\right]
$$

The KL divergence is a measure of how different two probability distributions are. It is always non-negative and is equal to zero if and only if the two distributions are the same almost everywhere.

Since $\log p(y_i \mid \theta)$ is constant with respect to $q_i$, within any set $Q$ of distributions on $\eta_i$ (with the same support as $p(\eta_i \mid y_i, \theta)$), finding the $q_i \in Q$ that maximizes an individual's ELBO is equivalent to minimizing its KL divergence with respect to the associated individual posterior $p(\eta_i \mid y_i, \theta)$. That is:

$$
\begin{aligned}
\arg\max_{q_i \in Q} \mathrm{ELBO}_i &= \arg\max_{q_i \in Q} \left( \underbrace{\log p(y_i \mid \theta)}_{\text{constant in } q_i} - \mathrm{KL}(q_i(\eta_i) \parallel p(\eta_i \mid y_i, \theta)) \right) \\
&= \arg\max_{q_i \in Q} (-\, \mathrm{KL}(q_i(\eta_i) \parallel p(\eta_i \mid y_i, \theta))) \\
&= \arg\min_{q_i \in Q} \mathrm{KL}(q_i(\eta_i) \parallel p(\eta_i \mid y_i, \theta))
\end{aligned}
$$

The KL divergence is minimized to zero when $q_i(\eta_i) = p(\eta_i \mid y_i, \theta)$. In this case, given Eq. 1, the ELBO equals the log marginal likelihood, i.e., an optimal lower bound. Unfortunately, in general, the individual posterior does not have a closed form, so in practice we have to approximate it with $q_i(\eta_i)$ instead.

### 2.2.3 Minorization Maximization (MM)

If we would like to maximize a function $f(\theta)$, but we cannot do so directly, one approach is to find a function $g(\theta \mid \theta')$ (i.e. a function of $\theta$ for a fixed $\theta'$, $g(\cdot \mid \theta') : \theta \mapsto g(\theta \mid \theta')$) that is a lower bound on $f(\theta)$ for all $\theta$ and is ideally tight at $\theta = \theta'$. Maximizing $g(\theta \mid \theta')$ with respect to $\theta$ may then also increase $f(\theta)$, since $g(\theta \mid \theta') \leq f(\theta)$ for all $\theta$. This is the main idea behind the MM algorithm (Lange et al., 2000). The function $g(\theta \mid \theta')$ is called a minorizer of $f(\theta)$ at $\theta'$. The MM algorithm iteratively constructs a minorizer at the current estimate of the parameters and then maximizes it to get an updated estimate of the parameters. This process is repeated until convergence.

In the context of maximizing the (log) marginal likelihood, one can construct an instance of the MM algorithm by alternating between:

1. Given $\theta$, find the $q_i$ that maximizes the ELBO (minorizer) for each subject.
2. Given $q_i$ for each subject, find the $\theta$ that maximizes the population ELBO.

### 2.3 MCEM and SAEM

Using the optimal proposal $q_i(\eta_i) = p(\eta_i \mid y_i, \theta)$, the expectation in the ELBO can be written as an integral over the posterior of the random effects.

$$\begin{aligned}
\text{ELBO}_i &= E_{q_i(\eta_i)}\left[\log \frac{p(y_i, \eta_i \mid \theta)}{q_i(\eta_i)}\right] \\
&= E_{p(\eta_i \mid y_i, \theta)}[\log p(y_i, \eta_i \mid \theta) - \log p(\eta_i \mid y_i, \theta)] \\
&= \int p(\eta_i \mid y_i, \theta)[\log p(y_i, \eta_i \mid \theta) - \log p(\eta_i \mid y_i, \theta)] \; d\eta_i \\
&= \int p(\eta_i \mid y_i, \theta)[\log p(y_i \mid \eta_i, \theta) + \log p(\eta_i \mid \theta) - \log p(\eta_i \mid y_i, \theta)] \; d\eta_i
\end{aligned}$$

In NLME models, this integral cannot be computed analytically. If it were possible to compute the above integral analytically, then integrating the random effects in the original marginal likelihood integral would have also been possible, and the problem of computing the marginal likelihood would have been solved.

MCEM addresses this problem by defining the ELBO with a different proposal distribution $q_i(\eta_i)$ which is the individual posterior from the previous iteration:

$$q_i(\eta_i) = p(\eta_i \mid y_i, \theta')$$

where $\theta'$ is the value of $\theta$ from the previous iteration. The individual ELBO can then be written as:

$$\begin{aligned}
\text{ELBO}_i &= E_{q_i(\eta_i)}\left[\log \frac{p(y_i, \eta_i \mid \theta)}{q_i(\eta_i)}\right] \\
&= E_{p(\eta_i \mid y_i, \theta')}\left[\log \frac{p(y_i, \eta_i \mid \theta)}{p(\eta_i \mid y_i, \theta')}\right] \\
&= E_{p(\eta_i \mid y_i, \theta')}[\log p(y_i, \eta_i \mid \theta)] - E_{p(\eta_i \mid y_i, \theta')}[\log p(\eta_i \mid y_i, \theta')] \\
&= \underbrace{E_{p(\eta_i \mid y_i, \theta')}[\log p(y_i \mid \eta_i, \theta) + \log p(\eta_i \mid \theta)]}_{\text{Q-function}} - \underbrace{E_{p(\eta_i \mid y_i, \theta')}[\log p(\eta_i \mid y_i, \theta')]}_{\text{constant in } \theta} \qquad (2)
\end{aligned}$$

The above expression is a valid lower bound on the log marginal likelihood because the support of $p(\eta_i \mid y_i, \theta')$ is the same as the support of the prior $p(\eta_i \mid \theta)$. Decoupling the sampling distribution $q_i(\eta_i) = p(\eta_i \mid y_i, \theta')$ from the parameters being estimated $\theta$ allows us to drop the generally intractable last term, when maximizing the ELBO with respect to $\theta$, since it is constant with respect to $\theta$ and only depends on the previous value of $\theta$ (i.e., $\theta'$).

The remaining expectation is commonly referred to as the *Q-function* in the EM literature. The Q-function is equal to the individual ELBO up to a constant and it can be approximated by a Monte Carlo average using samples from $p(\eta_i \mid y_i, \theta')$. To obtain these samples, MCMC can be used for each subject individually using the subject's observations $y_i$ and the previous value of $\theta$ (i.e., $\theta'$).

$$E_{p(\eta_i \mid y_i, \theta')}[\log p(y_i \mid \eta_i, \theta) + \log p(\eta_i \mid \theta)] \approx \frac{1}{M} \sum_{j=1}^{M} \big[\log p\big(y_i \mid \eta_i = \eta_{i,j}, \theta\big) + \log p\big(\eta_i = \eta_{i,j} \mid \theta\big)\big]$$

This gives rise to the MCEM algorithm (Wei & Tanner, 1990). MCEM alternates between the following two steps:

1. Given the previous value of $\theta = \theta'$, MCMC samples $\eta_{i,j}$ are obtained from each individual's posterior, then the above average is computed. This is known as the *expectation* step of the EM algorithm, a special instance of the *minorization* step in the MM framework.
2. Given the MCMC samples, the following estimator is maximized with respect to $\theta$. This is known as the *maximization* step in both the EM algorithm and the MM framework.

$$\frac{1}{M} \sum_{i=1}^{N} \sum_{j=1}^{M} \big(\log p\big(y_i \mid \eta_i = \eta_{i,j}, \theta\big) + \log p\big(\eta_i = \eta_{i,j} \mid \theta\big)\big)$$

MCEM is known to converge to a local maximum of the marginal likelihood *almost surely* (i.e., the set of realizations in which it does not converge has probability zero) (Wei & Tanner, 1990) under mild conditions, if the number of MCMC samples $M$ is increased at each iteration, eventually going to infinity as the number of iterations goes to infinity. In practice, a large but finite $M$ may be required to achieve good performance. This makes MCEM computationally expensive.

SAEM (Delyon et al., 1999; Kuhn & Lavielle, 2004; 2005) is a variant of MCEM that addresses this issue by using a stochastic approximation scheme to update the estimate of the Q-function at each iteration. The stochastic approximation is a weighted average scheme, effectively recycling previous MCMC samples with a decaying weight and combining them with the new MCMC samples obtained at the current iteration to update the estimate of the Q-function. The weights are chosen to decay over time, giving less weight to the samples in the early iterations and more weight to the samples in the later iterations. This allows SAEM to achieve good performance with a small number of MCMC samples per iteration, making it much more computationally efficient than MCEM.

SAEM is also known to converge *almost surely* to a local maximum of the marginal likelihood under mild conditions, if its step size is chosen appropriately and the number of iterations goes to infinity, even with a fixed number ($\geq 1$) of MCMC samples per iteration, in theory. In practice, ensuring that the samples are of good quality and that MCMC sampling is done correctly can be challenging. Additionally, picking a suitable step size schedule and number of iterations to terminate the algorithm can be difficult. Therefore, SAEM requires careful tuning of its hyper-parameters to achieve good performance in practice.

It is important to note that the Q-function which is the objective maximized by MCEM and SAEM (Eq. 2) is a constant away from the actual ELBO. So even if the ELBO is a perfect lower bound equal to the log marginal likelihood and $M$ is large, the complete ELBO is never actually calculated during MCEM and SAEM. Therefore, to obtain an estimate of the marginal likelihood after estimating $\theta$, a separate step (e.g., using IS) is required.

Different MCMC algorithms give rise to different variants of MCEM and SAEM. For example, one can use Hamiltonian Monte Carlo (HMC) (Neal, 2011) or one of its No-U-Turn Sampler (NUTS) variants (Hoffman & Gelman, 2014), which are more efficient than Metropolis-Hastings for sampling from high-dimensional distributions. This allows MCEM and SAEM to be applied to models with higher-dimensional random effects than would be possible with Metropolis-Hastings. To our knowledge, this pairing of HMC or NUTS with MCEM or SAEM has not been explored in the NLME literature; the closest attempt we found incorporates HMC into a stochastic EM scheme for *linear* mixed-effects models (Zakkour et al., 2023). There is nonetheless no theoretical obstacle: MCEM and SAEM place no restriction on the MCMC kernel used to sample the individual posteriors, so any valid kernel can be substituted, and the more favorable scaling of such kernels with the random-effect dimension (Beskos et al., 2013; Roberts et al., 1997) then carries over. However, MCMC is generally computationally expensive and can be difficult to tune and diagnose when convergence issues arise. Having to run MCMC for each subject in each iteration is a significant computational burden, especially when the number of subjects is large or the model is expensive to evaluate.

Alternatively, one can use VI to first approximate the individual posteriors and then optionally use the VI posterior approximation to guide other MCMC algorithms, e.g., as a proposal in a Metropolis-Hastings algorithm (Hastings, 1970; Metropolis et al., 1953). This was the approach taken by Freitas et al. (2001) and benchmarked in an NLME context by Karimi et al. (2020). A comprehensive empirical comparison of VEM against these and other NLME model fitting methods would be interesting, but is out of scope given the length of this paper and is left to future work. In the next section, we describe a related idea that is the basis of the VEM algorithm implemented in Pumas.

# 3 Variational Expectation Maximization (VEM)

The main difference between MCEM and VEM is that VEM, instead of using MCMC to sample from the individual posterior distributions $p(\eta_i \mid y_i, \theta)$, uses parametric distributions to approximate them; an approach known as VI (Jordan et al., 1999).

Figure 1 gives an overview of how VEM is derived and optimized; the remainder of this section develops each step in detail.

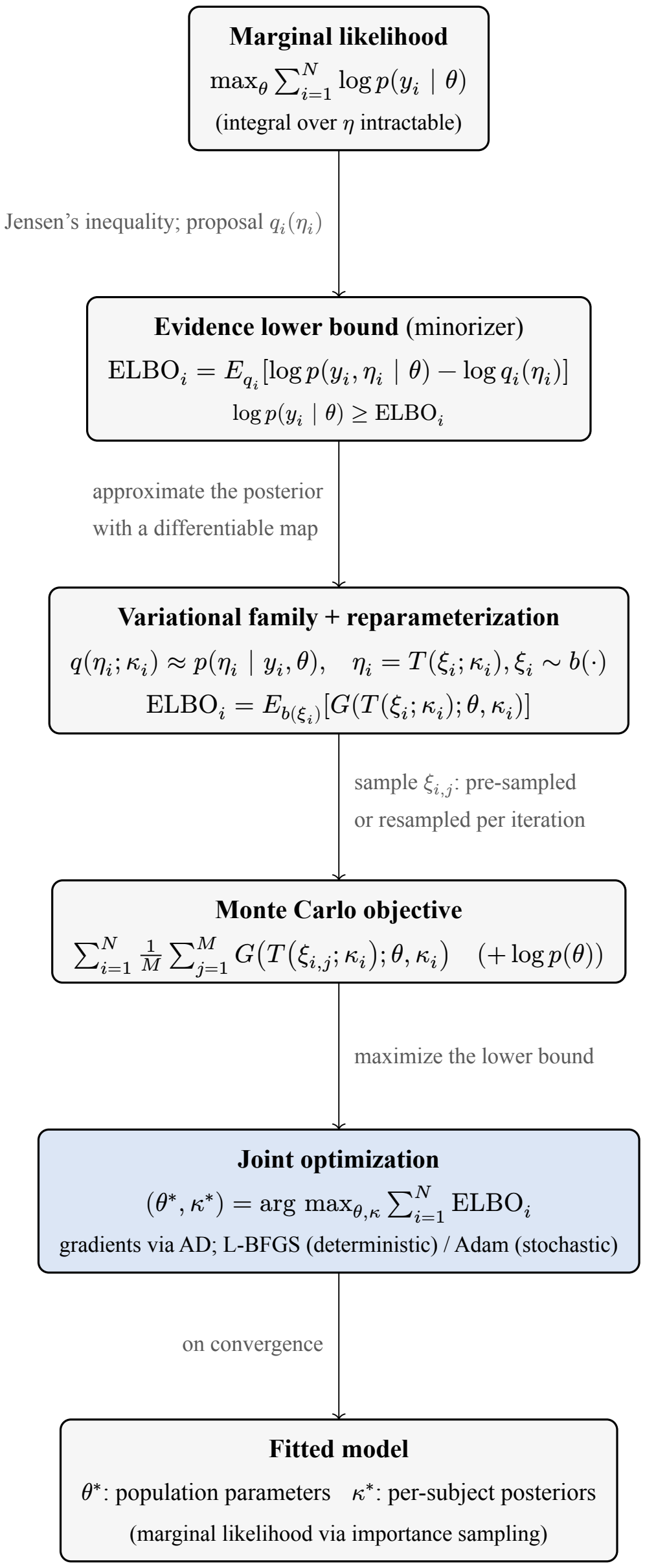

Figure 1: Overview of the VEM derivation and optimization loop. Starting from the intractable marginal likelihood, a proposal distribution and Jensen's inequality yield the ELBO, a lower bound (minorizer) on the log marginal likelihood. A parametric variational family together with the reparameterization trick makes the ELBO differentiable in its parameters, and a Monte Carlo average gives a tractable objective. Joint gradient-based optimization of the population parameters $\theta$ and the variational parameters $\kappa$ then maximizes this objective, iterating until convergence. Each step is developed in the remainder of this section.

### 3.1 From MM to VEM

We start with the MM algorithm (Section 2.2.3). Recall that the individual ELBO can be written as:

$$\text{ELBO}_i = E_{q_i(\eta_i)}[\log p(y_i \mid \eta_i, \theta) + \log p(\eta_i \mid \theta) - \log q_i(\eta_i)]$$

Our goal here is to use a parametric distribution $q(\eta_i; \kappa_i)$ as the proposal distribution for each individual, where $\kappa_i$ are its parameters. From this point on, we adopt the notation $q(\eta_i; \kappa_i)$ in place of $q_i(\eta_i)$ to make the parameter dependence explicit. What is shared across subjects here is the functional form of $q$, not the fitted distribution: each subject has its own parameter vector $\kappa_i$, so the fitted distributions generally differ from one subject to the next. With a Gaussian family, for instance, $\kappa_i$ collects subject $i$'s mean and covariance; with a mixture family it also collects the mixing ratio, so subjects can differ in the shape of the distribution and not only in its location and scale. For some specific parameter values, $q(\eta_i; \kappa_i)$ is known as a *variational distribution*, and the set of all possible values is known as a *variational family*. A good variational family is one that admits a variational distribution which approximates the individual posterior distribution $p(\eta_i \mid y_i, \theta)$ well, since we have established that this is the optimal proposal distribution for IS.

Given the above proposal distribution, the individual ELBO becomes:

$$\text{ELBO}_i = E_{q(\eta_i;\kappa_i)}\underbrace{[\log p(y_i \mid \eta_i, \theta) + \log p(\eta_i \mid \theta) - \log q(\eta_i; \kappa_i)]}_{G(\eta_i;\theta,\kappa_i)}$$

For brevity, calling the inner term $G(\eta_i; \theta, \kappa_i)$ we get:

$$\text{ELBO}_i = E_{q(\eta_i;\kappa_i)}[G(\eta_i; \theta, \kappa_i)]$$

We would like to use gradient-based optimization to find the $\kappa_i$ that maximizes the individual ELBO. The challenge here is that the distribution we are taking the expectation with respect to, $q(\eta_i; \kappa_i)$, is also parameterized by $\kappa_i$. Therefore, we cannot simply move the gradient operator inside the expectation and differentiate $G(\eta_i; \theta, \kappa_i)$ with respect to $\kappa_i$. Instead, we need to account for the fact that changing $\kappa_i$ also changes the distribution of $\eta_i$.

One classical way to differentiate such an expectation is the *score function estimator*, also known as the *likelihood ratio* or Likelihood Ratio Policy Gradient (REINFORCE) estimator (Glynn, 1990; Williams, 1992). Using the identity:

$$\nabla_{\kappa_i} q(\eta_i; \kappa_i) = q(\eta_i; \kappa_i) \nabla_{\kappa_i} \log q(\eta_i; \kappa_i),$$

the gradient of the individual ELBO can be written as an integral and differentiated under the integral sign:

$$\begin{aligned}
\nabla_{\kappa_i} \text{ELBO}_i &= \nabla_{\kappa_i} \int q(\eta_i; \kappa_i) G(\eta_i; \theta, \kappa_i) \, \mathrm{d}\eta_i \\
&= \int \nabla_{\kappa_i} [q(\eta_i; \kappa_i) G(\eta_i; \theta, \kappa_i)] \, \mathrm{d}\eta_i \\
&= \int \left[ G(\eta_i; \theta, \kappa_i) \nabla_{\kappa_i} q(\eta_i; \kappa_i) + q(\eta_i; \kappa_i) \nabla_{\kappa_i} G(\eta_i; \theta, \kappa_i) \right] \mathrm{d}\eta_i \\
&= \int q(\eta_i; \kappa_i) \left[ G(\eta_i; \theta, \kappa_i) \nabla_{\kappa_i} \log q(\eta_i; \kappa_i) + \nabla_{\kappa_i} G(\eta_i; \theta, \kappa_i) \right] \mathrm{d}\eta_i
\end{aligned}$$

where the third equality is the product rule and the fourth substitutes the identity above. Recognizing the integral as an expectation with respect to $q(\eta_i; \kappa_i)$, this is

$$\begin{aligned}
\nabla_{\kappa_i} \mathrm{ELBO}_i &= E_{q(\eta_i;\kappa_i)}\left[G(\eta_i;\theta,\kappa_i)\nabla_{\kappa_i} \log q(\eta_i;\kappa_i) + \nabla_{\kappa_i} G(\eta_i;\theta,\kappa_i)\right] \\
&= E_{q(\eta_i;\kappa_i)}\left[G(\eta_i;\theta,\kappa_i)\nabla_{\kappa_i} \log q(\eta_i;\kappa_i)\right] + E_{q(\eta_i;\kappa_i)}\left[\nabla_{\kappa_i} G(\eta_i;\theta,\kappa_i)\right] \\
&= E_{q(\eta_i;\kappa_i)}\left[G(\eta_i;\theta,\kappa_i)\nabla_{\kappa_i} \log q(\eta_i;\kappa_i)\right] - \underbrace{E_{q(\eta_i;\kappa_i)}\left[\nabla_{\kappa_i} \log q(\eta_i;\kappa_i)\right]}_{\text{expected score is 0}} \\
&= E_{q(\eta_i;\kappa_i)}\left[G(\eta_i;\theta,\kappa_i)\nabla_{\kappa_i} \log q(\eta_i;\kappa_i)\right]
\end{aligned}$$

which only requires sampling from $q(\eta_i; \kappa_i)$ and evaluating its score $\nabla_{\kappa_i} \log q(\eta_i; \kappa_i)$. This makes it applicable to a broad class of variational families, including discrete ones, and is the basis of black-box variational inference (Ranganath et al., 2014). Its main practical drawback is that the resulting Monte Carlo gradient estimator tends to exhibit high variance (Ranganath et al., 2014).

When the variational family is continuous and differentiable in $\kappa_i$, a lower-variance alternative is the *reparameterization trick* (Kingma & Welling, 2013; Rezende et al., 2014), which is the approach we adopt in VEM. The idea is that a parametric probability distribution can be written as a deterministic parametric function of a fixed base distribution that does not depend on $\kappa_i$. That is:

- We define a base random variable $\xi_i$, with the same dimensionality as $\eta_i$, but with a probability distribution $b(\xi_i)$ that has fixed parameters, e.g. $N(0, I)$.
- We define a parametric invertible (for reasons that will be clear later) transformation $T(\xi_i; \kappa_i)$ whose codomain is the support of $\eta_i$.

To sample from this proposal distribution for a specific subject, we first draw a sample of $\xi_i$ from its distribution:

$$\xi_i \sim b(\cdot)$$

Then, for the given parameter values, we evaluate $T(\xi_i; \kappa_i)$ to obtain a sample of $\eta_i$:

$$\eta_i = T(\xi_i; \kappa_i)$$

This way, we can rewrite each $\mathrm{ELBO}_i$ as an expectation with respect to a probability distribution that does not depend on $\kappa_i$:

$$\begin{aligned}
\mathrm{ELBO}_i &= E_{q(\eta_i;\kappa_i)}[G(\eta_i;\theta,\kappa_i)] \\
&= E_{b(\xi_i)}[G(T(\xi_i;\kappa_i);\theta,\kappa_i)]
\end{aligned}$$

### 3.2 Density evaluation of variational families

Computing $G(T(\xi_i; \kappa_i); \theta, \kappa_i)$ requires us to evaluate the PDF $q(\eta_i; \kappa_i)$. If we choose a base random variable for which density evaluation is easy to perform, we can evaluate $q(\eta_i; \kappa_i)$ by using the change of variables formula:

$$q(\eta_i; \kappa_i) = \frac{b(\xi_i)}{\left|\frac{\partial T}{\partial \xi_i}(\xi_i; \kappa_i)\right|}$$

$$\xi_i = T^{-1}(\eta_i; \kappa_i)$$

where $T^{-1}$ is the inverse of the transformation $T$ and $|\frac{\partial T}{\partial \xi_i}|$ is the absolute value of the determinant of the Jacobian of the transformation $T$ with respect to $\xi_i$.

### 3.3 The Gaussian family

The simplest variational family is the Gaussian family, where $b(\xi_i)$ is a standard multivariate Gaussian distribution and $T(\xi_i; \kappa_i)$ is an affine transformation. With $L_i$ being a lower triangular matrix (in Pumas it is diagonal by default, but optionally dense), and $\mu_i$ a column vector:

$$\kappa_i = (L_i, \mu_i)$$

$$\xi_i \sim N(0, I)$$

$$\eta_i = T(\xi_i; \kappa_i) = L_i \cdot \xi_i + \mu_i \tag{3}$$

$T$ can also optionally include an additional invertible transformation $D$:

$$T(\xi_i; \kappa_i) = D(L_i \cdot \xi_i + \mu_i; \kappa_i) \tag{4}$$

Note that in Pumas, VEM supports non-Gaussian and constrained random effects. Therefore, before approximating the posterior with a Gaussian variational distribution, we first transform the random effects using a domain transformation if the random effect's distribution is constrained, e.g. log transforming a log-normally distributed random effect. If the resulting random effects after the domain transformation are Gaussian (e.g. they were log-normal or logit-normal previously), they are then standardized by their mean and covariance. The transformed random effects' posterior is then the one that is approximated by a Gaussian variational distribution, parameterized as shown in Eq. 3. This is mathematically equivalent to including de-standardization and the inverse domain transformation as part of the invertible function $D$ in Eq. 4 to ensure that the support of $q(\eta_i; \kappa_i)$ is the same as the support of the prior $p(\eta_i \mid \theta)$. So for the rest of this paper, assume that either $T$ includes any necessary de-standardization and inverse domain transformation, or that $\eta_i$ refers to the transformed and sometimes standardized random effects.

### 3.4 Expectation revisited

In VEM, the individual ELBO can be expressed as:

$$\text{ELBO}_i = E_{b(\xi_i)}[G(T(\xi_i; \kappa_i); \theta, \kappa_i)]$$

To maximize the ELBO, the above expectation is approximated by sampling $\xi_i$ for each subject $i$.

$$\mathrm{ELBO}_i \approx \frac{1}{M} \sum_{j=1}^{M} G\big(T(\xi_{i,j}; \kappa_i); \theta, \kappa_i\big) \tag{5}$$

where $\{\xi_{i,1}, \xi_{i,2}, ..., \xi_{i,M}\}$ are $M$ samples from the base distribution $b(\xi_i)$ for each subject $i$.

When optimizing the above approximation, we can use one of the following approaches:

1. Resample $M$ values of $\xi_i$ from $b(\xi_i)$ in every optimization iteration for each subject $i$. This yields a stochastic objective that requires stochastic optimization algorithms, such as stochastic gradient descent (SGD) or Adam (Adaptive Moment Estimation) (Kingma & Ba, 2015). This estimator of the ELBO is unbiased but typically leads to high gradient variance and slow convergence of the optimization.
2. Pre-sample a fixed, sufficiently large set of samples for $\xi_i$ for each subject, and then use a deterministic optimization algorithm. This estimator is deterministic (zero variance during the optimization) but biased. However, if the sample size $M$ is large enough, the bias can become small enough to have little effect on the optimization.

The deterministic-objective approach has also been recently studied in the variational inference literature, under the name *sample average approximation* (Burroni et al., 2024; Giordano et al., 2024), where it has been shown to enable the use of quasi-Newton optimizers and line search, yielding faster and more reliable convergence than stochastic gradient methods. We had the same motivation and found the same benefits of using a deterministic objective for optimization, which is why deterministic VEM is the default variant of VEM in Pumas.

Note that with the deterministic objective, the quantity actually optimized is the fixed finite-sample Monte Carlo estimator in Eq. 5, not the exact ELBO. The ELBO is a lower bound on the log marginal likelihood, but this finite-sample estimator is not necessarily so: averaging $G$ over a fixed set of $M$ samples of $\xi_i$ introduces sampling error, so the estimator can fall above the exact ELBO, and hence occasionally above the log marginal likelihood that the ELBO bounds.

### 3.5 Joint optimization

In the minorization step, the goal is to find the best (highest) lower bound on the log marginal likelihood for each individual:

$$\kappa_i^* = \arg\max_{\kappa_i} \mathrm{ELBO}_i$$

Since the objectives for all the subjects are separate, doing the above optimization for each subject $i$ is equivalent to doing a joint maximization:

$$\kappa^* = \arg\max_{\kappa} \sum_{i=1}^{N} \mathrm{ELBO}_i$$

where $\kappa = (\kappa_1, ..., \kappa_N)$ is the vector of all the variational distribution parameters for all subjects.

Similarly, the maximization step tries to find the $\theta$ that maximizes the population ELBO:

$$\theta^* = \arg\max_{\theta} \sum_{i=1}^{N} \mathrm{ELBO}_i$$

The original VEM algorithm (Jordan et al., 1999) alternates between the minorization and maximization steps, but since they have the same objective, both optimizations can be done jointly as follows:

$$(\theta^*, \kappa^*) = \arg\max_{\theta, \kappa} \sum_{i=1}^{N} \mathrm{ELBO}_i$$

Gradient-based optimization algorithms can be used to perform this optimization. Joint optimization was also adopted in the VAE literature (Kingma & Welling, 2013; Rezende et al., 2014) which uses a variant of the VEM algorithm. A more detailed discussion of the connection between VEM and VAE can be found in Section 3.10.

### 3.6 Mini-batch VEM

When using the stochastic estimator, we can further sub-sample the population used to construct the population ELBO, if the population size is large. In each optimization iteration, a random $s\%$ of the population is selected. Denote the indices of the selected subjects in the $k^{\text{th}}$ optimization iteration by $S_k$. The population ELBO optimized in iteration $k$ is:

$$\frac{100}{s} \cdot \sum_{i \in S_k} \mathrm{ELBO}_i$$

Such a *mini-batch* approach can lead to faster convergence if the population size is too large, and is similar to the mini-batch stochastic gradient descent (SGD) approach commonly used in training neural networks.

The $100/s$ scalar is important to ensure that the expected values of the objective and its gradient are the same as those of the full data case, regardless of batch size. This is especially important if we incorporate a prior distribution on the population parameters as additional regularization (Section 3.7).

These choices together define the three variants of VEM implemented in Pumas: deterministic VEM (pre-sampled $\xi_i$, full population), stochastic VEM (resampled $\xi_i$ per iteration, full population), and stochastic VEM with population sub-sampling. Running a few iterations of the sub-sampled variant is particularly useful when fitting complex NLME models to a very large population: it can find good parameter estimates quickly, which can then be used as a starting point in a deterministic VEM fit with the full population.

Currently, the sub-sample $S_k$ is drawn uniformly at random from the population in each iteration. An alternative is *stratified* sub-sampling, in which the population is partitioned by relevant covariates (e.g., dose, body weight, or disease status) and each mini-batch is drawn so that its covariate composition matches that of the full population. Matching the mini-batch characteristics to the population in this way could reduce the variance of the sub-sampled objective and its gradient, potentially improving convergence. Stratified sub-sampling is not implemented in Pumas and is left as future work.

### 3.7 Maximum A Posteriori (MAP) estimation of the population parameters

In some cases, if a model is over-parameterized or the data is limited, the population parameters $\theta$ may not be identifiable given the data available. This is especially true for NLME models that incorporate a neural network with many parameters. In such over-parameterized cases, a prior distribution on the population parameters can be used to regularize the parameter estimation. In the case of neural-network-based models, the prior on each parameter of the neural network is typically a Gaussian distribution with zero mean and a constant variance hyperparameter, controlling the regularization strength. In the case of pediatric pharmacokinetic models, the prior may be derived from the posterior distribution of the parameters given data from a similar study on an adult population, after performing allometric scaling of the model. In Pumas, when a prior $p(\theta)$ is used on the population parameters, marginal MAP estimation can be performed instead of just maximizing the marginal likelihood. The objective function becomes:

$$\theta^* = \arg\max_{\theta} p(y \mid \theta) \cdot p(\theta)$$

where $p(y \mid \theta)$ is the population marginal likelihood. The same VEM algorithm described above can be used to optimize this objective, with the log prior added to the population ELBO as follows:

$$\log p(\theta) + \sum_{i=1}^{N} \text{ELBO}_i$$

When sub-sampling the population, the objective in iteration $k$ becomes:

$$\log p(\theta) + \frac{100}{s} \cdot \sum_{i \in S_k} \text{ELBO}_i$$

### 3.8 Optimization algorithms

In Pumas, deterministic VEM uses the Limited-memory Broyden-Fletcher-Goldfarb-Shanno (L-BFGS) optimization algorithm (Nocedal & Wright, 2006), which is a quasi-Newton method that approximates the inverse Hessian of the objective function using its historical gradients to calculate the search direction in each optimization iteration. A backtracking line search is then performed to find the best solution along the search direction. Note that for large NLME models and/or large datasets, it is important to use L-BFGS and not the original Broyden-Fletcher-Goldfarb-Shanno (BFGS) algorithm, because BFGS maintains a dense approximation of the inverse Hessian whose memory and time requirements grow quadratically with the total number of optimized parameters, as quantified below. Let $p$ be the number of population parameters, $v$ be the number of parameters in the variational distribution for each subject, $N$ be the number of subjects in the dataset, $r$ be the number of random effects, and $l$ be the limited memory size in L-BFGS. Using a Gaussian variational family with diagonal covariance, $v$ scales linearly with $r$ and therefore the time and memory requirements of BFGS scale with $O\big((p+N\cdot v)^2\big)=O\big((p+N\cdot r)^2\big)$ while the time and memory requirements of L-BFGS scale with $O(l\cdot(p+N\cdot v))$. L-BFGS will therefore be faster as long as $l \ll p+N\cdot v$.

The stochastic VEM uses the Adam (Kingma & Ba, 2015) optimization algorithm, which is a variant of stochastic gradient descent that adapts the learning rate for each parameter based on the first and second moments of the gradients. Additionally, in Pumas, gradient scaling and gradual learning rate decay are also performed for numerical stability. By default, the stochastic VEM in Pumas uses the entire population, without sub-sampling.

### 3.9 ELBO gradient

Both the L-BFGS and Adam optimization algorithms require the gradient of the ELBO with respect to both $\theta$ and $\kappa$. The gradient of the population ELBO is the sum of the gradients of the individual ELBOs. The gradient computation involves applying the chain rule to decompose the ELBO into terms that can be differentiated using AD. The choice of AD mode is dictated by the relative size of inputs and outputs: for a scalar ELBO with a large total number of population parameters and random effects, reverse-mode AD computes the gradient at a cost that is bounded by a small constant multiple of one ELBO evaluation, while forward-mode AD's cost grows linearly with the number of inputs. For small models, the lower overhead of forward-mode AD typically makes it faster despite this asymptotic disadvantage. When the NLME model includes an ODE, AD is applied directly to the discretized (numerical) solution of the ODE (known as the *discrete sensitivity* method in forward mode and the *discrete adjoint* method in reverse mode), which guarantees gradients that are consistent with the discretized solution actually used to evaluate the log likelihood and avoids the reverse-time stability issues of the continuous adjoint method.

In Pumas, VEM is implemented using both forward-mode and reverse-mode AD, while Laplace, FOCE, and FO are implemented using only forward-mode AD. See Appendix A.1 for the full ELBO gradient derivation and Appendix A.2 for a more detailed discussion of AD techniques and their interaction with ODE solvers.

### 3.10 Amortized inference and connection to VAEs and PPCA

In this section, we discuss the connection between the VEM algorithm and VAEs and PPCA. VAEs are often interpreted as a probabilistic generalization of autoencoders. The original (non-variational) autoencoder model (Hinton & Salakhutdinov, 2006) is made of two Neural Networks (NNs):

1. The first NN learns to encode the input data $y$ into a lower-dimensional latent space $z = f_{\text{encoder}}(y;\kappa)$, and
2. The second NN decodes $z$ back to the original data space $\hat{y} = f_{\text{decoder}}(z;\theta)$

where $\kappa$ and $\theta$ are the parameters of the encoder and decoder functions, respectively. The goal of an autoencoder is to learn the parameters $(\theta,\kappa)$ that minimize the expectation of some cost function $C$, representing the *reconstruction error*, under the data distribution:

$$\begin{aligned}(\theta^*,\kappa^*) &= \arg\min_{\theta,\kappa} E_{p_{\text{true}}(y)}[C(y,\hat{y}(y;\theta,\kappa))] \\ &= \arg\min_{\theta,\kappa} E_{p_{\text{true}}(y)}[C(y,f_{\text{decoder}}(f_{\text{encoder}}(y;\kappa);\theta))]\end{aligned}$$

The autoencoder is akin to a nonlinear Principal Component Analysis (PCA) (Hotelling, 1933; Pearson, 1901), generalizing PCA. The encoded data $f_{\text{encoder}}(y;\kappa)$ plays the role of the projection of the input data onto the principal components $V^T \cdot y$ in classical PCA, where $V$ is the matrix whose columns are the top principal components of the data. In an autoencoder, the decoder is defined by a nonlinear function (neural network), while in PCA, the decoder is $V \cdot z$, which is a linear function. Therefore, the principal components (and their transpose) from classical PCA can be seen as a non-unique but globally optimal linear autoencoder, where the encoder and decoder are linear functions, the cost function is the mean squared error (MSE), and the data is centered.

The VAE extends the autoencoder by introducing:

1. A prior distribution on the latent variables $p(z)$,
2. A probability distribution for $y \mid z$ parameterized by the output of the decoder, $f_{\text{decoder}}(z;\theta)$; and
3. A distribution $q(z; y, \kappa)$ given each data point $y$, parameterized by the output of the encoder, $f_{\text{encoder}}(y;\kappa)$.

where the distribution parameterized by the encoder's output $f_{\text{encoder}}(y;\kappa)$ approximates the posterior distribution $p(z \mid y, \theta)$.

The latent variables and decoder in the VAE generalize the PPCA model (Tipping & Bishop, 1999):

$$z \sim N(0, I)$$
$$y \sim N(W \cdot z + \mu, \sigma^2 \cdot I)$$

where the parameters of the decoder in a PPCA are $\theta = (W, \mu, \sigma)$. In PPCA, the posterior distribution $p(z \mid y, \theta)$ has a closed form solution:

$$p(z \mid y, \theta) = N(M^{-1} \cdot W^T \cdot (y - \mu), \sigma^2 \cdot M^{-1})$$

$$M = W^T \cdot W + \sigma^2 \cdot I$$

Therefore, there is no need to approximate the posterior distribution in this case.

In VAEs, however, the posterior distribution $p(z \mid y, \theta)$ is generally not known in closed form. Therefore, we use a variational distribution $q_i(z)$ for each data point (subject) $y_i$ to approximate the posterior. However, the VAE approach does not parameterize each subject's variational distribution independently using a separate $\kappa_i$, as we proposed when presenting the VEM algorithm in this paper. This is because VAEs were designed to work with large datasets, where it is arguably inefficient to use different parameters for each subject/image/point. Instead, VAEs try to learn a function that maps each subject's observations $y_i$ to its posterior distribution $p(z \mid y, \theta)$. If we assume a Gaussian posterior approximation, the encoder function $f_{\text{encoder}}(y;\kappa)$ can output the mean and covariance matrix of the Gaussian approximation of the individual posterior for any given $y$. This idea of learning a mapping allows parameter sharing between all subjects' variational distributions, decoupling the number of parameters in $\kappa$ from the number of individuals in the dataset. This concept is commonly known as *amortized inference* (Cremer et al., 2018).

However, amortized inference introduces an additional layer of complexity and increases the potential for bias. First, the encoder needs to take an input $y$ which can have varying length, e.g., different subjects have a different number of observations. The posterior distribution may not be a simple transformation of the data $y$. This means that if the architecture and size of the encoder are not sufficient to fit the relationship between $y$ and $p(z \mid y, \theta)$, the encoder will under-fit. If instead the encoder is over-parameterized, it may over-fit the training data, failing to generalize to unseen $y$, or it may require many samples in the expectation step to fit properly. Despite all these additional complications, amortized inference is a powerful technique that allows VAEs to scale to large datasets and complex models. After training the VAE, the encoder can be used to quickly approximate the posterior distribution for new data points without the need to perform any additional VI optimization or MCMC. The connection of VAEs to amortized inference and the potential for bias due to *sub-optimality* when fitting the encoder was discussed in Cremer et al. (2018).

Another way to view a VAE is as an NLME model with:

1. Random effects $z \sim p(\cdot)$,
2. A neural network structural model, and
3. Population parameters $\theta$, representing the parameters of the decoder neural network.

$$z \sim p(\cdot)$$
$$y \sim p(\cdot \mid z, \theta)$$

Replacing $z$ with $\eta$ and allowing the prior $p(z)$ to depend on the population parameters $\theta$, e.g., to explicitly model the between-subject variability $\Omega$ as part of $\theta$, we get:

$$\eta \sim p(\cdot \mid \theta)$$
$$y \sim p(\cdot \mid \eta, \theta)$$

This is the canonical form of an NLME model. For most NLME analyses in pharmacometrics, however, the dataset is not large enough to justify amortized inference, so we do not use amortized inference in the VEM algorithm proposed in this paper. For a variant of the VEM algorithm that uses amortized inference with NLME models, see Arruda et al. (2024).

### 3.11 Connection to the Laplace method

In this section, we discuss the connection between the VEM algorithm and the Laplace method. One way to view maximizing the Laplace approximation of the marginal likelihood is as an approximation of the VEM algorithm. Let the variational family for an individual $i$ be $N(\mu_i, \Sigma_i)$ where $\kappa_i = (\mu_i, \Sigma_i)$ is the set of parameters of the variational family.

The individual ELBO can be written as:

$$\text{ELBO}_i = E_{q(\eta_i;\kappa_i)}[\log p(y_i \mid \eta_i, \theta) + \log p(\eta_i \mid \theta) - \log q(\eta_i; \kappa_i)]$$

$$= E_{q(\eta_i;\kappa_i)}[\log p(y_i, \eta_i \mid \theta)] \underbrace{- E_{q(\eta_i;\kappa_i)}[\log q(\eta_i;\kappa_i)]}_{H[q(\eta_i;\kappa_i)]}$$

$$= E_{q(\eta_i;\kappa_i)}[\log p(y_i, \eta_i \mid \theta)] + H[q(\eta_i;\kappa_i)] \tag{6}$$

where $H[q(\eta_i;\kappa_i)]$ is known as the *entropy* of the variational distribution $q(\eta_i;\kappa_i)$. For a multivariate Gaussian distribution, the entropy is given by:

$$H[q(\eta_i;\kappa_i)] = \frac{r}{2}\log(2\pi e) + \frac{1}{2}\log|\Sigma_i|$$

where $r$ is the number of random effects.

Next, we approximate $\log p(y_i, \eta_i \mid \theta)$ with a second-order Taylor expansion around its mode:

$$\eta_i^* = \arg\max_{\eta_i} p(y_i, \eta_i \mid \theta) = \arg\max_{\eta_i} p(\eta_i \mid y_i, \theta)$$

$$\log p(y_i, \eta_i \mid \theta) \approx \log p(y_i, \eta_i^* \mid \theta) + \frac{1}{2} \cdot (\eta_i - \eta_i^*)^T \cdot g''(\eta_i^*) \cdot (\eta_i - \eta_i^*)$$

where $g''(\eta_i^*)$ is the Hessian matrix at the mode, which is negative semi-definite and is assumed to be negative definite (i.e., the mode is non-degenerate) for the subsequent derivation.

$$g''(\eta_i^*) = \left.\frac{\partial^2 \log p(y_i, \eta_i \mid \theta)}{\partial \eta_i \, \partial \eta_i^T}\right|_{\eta_i = \eta_i^*}$$

Denote $g''(\eta_i^*)$ by $H_i$. We then plug in the approximation of $\log p(y_i, \eta_i \mid \theta)$ into the first term of Eq. 6, which has an analytic expectation after the approximation:

$$E_{q(\eta_i;\kappa_i)}[\log p(y_i, \eta_i \mid \theta)] \approx \log p(y_i, \eta_i^* \mid \theta) + \frac{1}{2} \cdot \mathrm{Tr}(H_i \cdot \Sigma_i) + \frac{1}{2} \cdot (\mu_i - \eta_i^*)^T \cdot H_i \cdot (\mu_i - \eta_i^*)$$

So the individual ELBO becomes:

$$\mathrm{ELBO}_i \approx \log p(y_i, \eta_i^* \mid \theta) + \frac{1}{2} \cdot \mathrm{Tr}(H_i \cdot \Sigma_i) + \frac{1}{2} \cdot (\mu_i - \eta_i^*)^T \cdot H_i \cdot (\mu_i - \eta_i^*) + \frac{r}{2} \cdot \log(2\pi e) + \frac{1}{2} \cdot \log|\Sigma_i|$$

In the minorization step, we want to find the $(\mu_i, \Sigma_i)$ that maximize the approximation of the individual ELBO.

$$(\mu_i^*, \Sigma_i^*) = \arg\max_{\mu_i, \Sigma_i} \mathrm{ELBO}_i$$

Note that under the second order approximation, the objective is separable so we can optimize for $\mu_i$ and $\Sigma_i$ separately. We start by optimizing for $\Sigma_i$. After dropping terms constant in $\Sigma_i$ and the common $1/2$ factor (neither of which affects the argmax):

$$\Sigma_i^* = \arg\max_{\Sigma_i}(\mathrm{Tr}(H_i \cdot \Sigma_i) + \log|\Sigma_i|)$$

The gradient of the above objective with respect to $\Sigma_i$ is:

$$H_i + \Sigma_i^{-1}$$

Setting the gradient to zero gives us the optimal covariance matrix:

$$\Sigma_i^* = -H_i^{-1}$$

Next, we optimize for $\mu_i$:

$$\mu_i^* = \arg\max_{\mu_i}\Big((\mu_i - \eta_i^*)^T \cdot H_i \cdot (\mu_i - \eta_i^*)\Big)$$

Since $H_i$ is negative definite, the above objective is maximized when $\mu_i^* = \eta_i^*$.

Plugging in the optimal variational distribution into the individual ELBO gives us the Laplace approximation of the marginal likelihood for subject $i$:

$$\begin{aligned}
\mathrm{ELBO}_i &\approx \log p(y_i, \eta_i^* \mid \theta) - \frac{r}{2} + \frac{r}{2}\log(2\pi e) - \frac{1}{2} \cdot \log|-H_i| \\
&= \log p(y_i, \eta_i^* \mid \theta) + \frac{r}{2}\log(2\pi) - \frac{1}{2} \cdot \log|-H_i| \\
\exp(\mathrm{ELBO}_i) &\approx p(y_i, \eta_i^* \mid \theta) \cdot \sqrt{\frac{(2\pi)^r}{|-H_i|}}
\end{aligned}$$

In other words, the Laplace method is an approximation of the VEM algorithm, where the log joint probability $\log p(y_i, \eta_i \mid \theta)$ is approximated with a second-order Taylor expansion around its mode $\eta_i^*$. This connection between the Laplace method and VI was noted by Friston et al. (2007), but the result was also implied by Kass & Raftery (1995) which Friston et al. (2007) cites.

The above result means that if the quadratic approximation happens to be a good one, the VEM algorithm will naturally converge to maximizing the Laplace approximation of the marginal likelihood. This also means that the VEM algorithm is asymptotically consistent under the same conditions as the Laplace method, which is when a single mode dominates the posterior distribution of the random effects $\eta_i$ as the number of observations goes to infinity.

However, unlike the Laplace method, the general VEM makes no assumptions about the smoothness of the posterior, its twice-differentiability, or even the existence of a dominant mode. This makes it more generally applicable in cases where the assumptions of the Laplace method do not hold. Of course, if these assumptions are not satisfied, a Gaussian variational distribution may not be a good approximation of the individual posterior, so one may need to use more complex variational distributions. Another attractive feature of the VEM algorithm is that it only requires a single level of differentiation to optimize the ELBO. By comparison, the Laplace method requires three levels of differentiation, and the FOCE (Lindstrom & Bates, 1990) and FO (Sheiner & Beal, 1980) methods require two.

# 4 Related Work

Variational methods have recently attracted independent interest in pharmacometrics, and it is useful to position the VEM variant presented in this paper relative to them. Figure 2 organizes these approaches by whether they use amortized inference. In amortized inference, a single neural network (an encoder) maps each subject's data to the parameters of that subject's variational distribution, so the per-subject variational parameters are produced by a shared, globally-fitted function rather than optimized independently for each subject. The variant presented in this paper is non-amortized: each subject has its own variational parameters, optimized jointly with the population parameters.

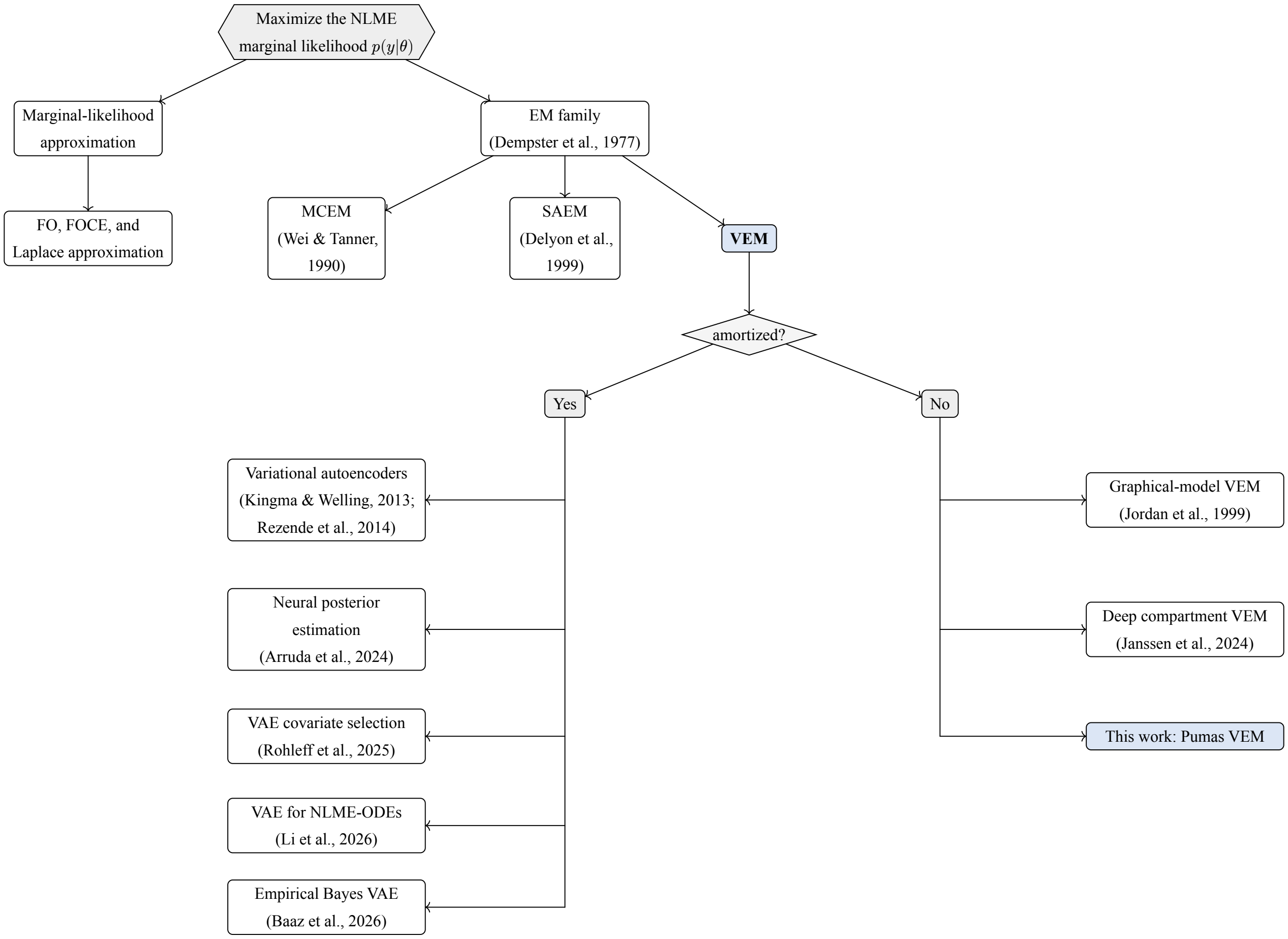

Figure 2: "Family tree" of methods for maximizing the NLME marginal likelihood. Arrows point from a general method to its specializations. VEM variants split by whether they use amortized inference. VEM and the Pumas variant presented in this work are shaded.

Arruda et al. (2024) introduce an amortized approach to NLME modeling based on neural posterior estimation, placing it on the amortized branch of Figure 2 alongside the VAE (Kingma & Welling, 2013; Rezende et al., 2014). The two main differences between Arruda et al. (2024) and the approach taken here are that our default variant uses non-stochastic (pre-sampled) sampling in the expectation step (Section 3.4) and that we do not use amortized inference (Section 3.10). Amortization replaces per-subject optimization with a single encoder network, which can accelerate inference on new subjects but ties the quality of each subject's posterior to the capacity and training of the shared encoder.

Janssen et al. (2024) estimate mixed effects in deep compartment models and compare FO and FOCE against a VI approach, reporting that VI is more reliable than FOCE on their examples. Their approach is non-amortized and per-subject, and in that respect it is on the same branch of Figure 2 as the variant presented here: it attaches a separate Gaussian variational distribution to each subject and optimizes its parameters jointly with the fixed effects, while treating the population parameters as point estimates, as we do. The main algorithmic difference is that they draw fresh Monte Carlo samples from the variational distribution at each gradient step and rely on the path-derivative gradient estimator (Roeder et al., 2017) to keep the variance of these stochastic gradients manageable, whereas our default variant pre-samples once and optimizes a deterministic objective (Section 3.4). One aspect of their comparison is worth noting when interpreting it. Their FOCE baseline optimizes the population parameters with Adam, a stochastic optimizer, whereas FOCE is conventionally paired with deterministic quasi-Newton or trust-region methods for that outer optimization; this difference may contribute to the unstable outer-optimization behavior they report. The FOCE implementation used in this paper is a mature and widely used one that, in most of the cases we have tested, agrees closely with VEM (Section 5.1), provided the model is not too large and the objective has no discontinuities that break the inner mode estimation of the Empirical Bayes Estimates (EBEs). Their results and ours are therefore best read as complementary evidence: the variational method of Janssen et al. (2024) is conceptually very similar to the VEM variant presented here, differing mainly in the sampling and optimization method, and our experiments add a reference point for how FOCE and VEM relate when FOCE is optimized in the conventional way.

Three further works share the encoder-based strategy of the VAE and Arruda et al. (2024) rather than the per-subject strategy adopted here, and so belong on the amortized branch of Figure 2. Rohleff et al. (2025) use a VAE for simultaneous parameter estimation and covariate selection, the latter being a task that is conceptually separate from the VEM estimation algorithm underlying the VAE. Li et al. (2026) apply amortized VEM to NLME models whose structural component is a system of ODEs. Baaz et al. (2026) propose a VAE in which the encoder is a transformer and the decoder is a neural ODE.

A broader aim of this paper is to clearly describe the VEM algorithm and its connections to related methods (the EM family, the VAE, PPCA, and the Laplace method), so that it is clear that VEM is an estimation algorithm that is separate from the model it is applied to. In the literature, VEM is frequently discussed together with "neural networks", "machine learning", and "artificial intelligence (AI)". We think it is worth separating these, because the association can obscure the generality of the VEM algorithm, which applies to any NLME model without requiring a neural network either in the model itself or, through amortization, in the estimation method. To be clear, VEM can reasonably be regarded as machine learning, and by extension artificial intelligence, in the same broad sense that its origin in PGMs (Jordan et al., 1999) places it within machine learning and machine learning is a part of artificial intelligence; by that same taxonomy, however, essentially all of NLME modeling would equally qualify as machine learning. When VEM is specifically labeled machine learning or artificial intelligence, it is usually because a neural network is involved, which we stress is an optional component rather than an intrinsic part of the algorithm. The association is reinforced when only the amortized variant of VEM is presented, since the encoder there is a neural network (Arruda et al., 2024; Baaz et al., 2026; Li et al., 2026; Rohleff et al., 2025). It is also reinforced when estimation methods are compared only within neural-network-based models. Our own scalability experiment (Section 5.2), Janssen et al. (2024), and Baaz et al. (2026) all use a neural network in the model, but we stress that this is incidental to the estimation algorithm being evaluated. In our case, the model simply serves as a convenient example of a large NLME model. We additionally apply VEM to other, more traditional model types: a standard Gaussian NLME model (Section 5.1) and a discrete-response Item Response Theory (IRT) model (Appendix A.4) with known true parameters. This makes it clear that the algorithm is independent of both the response distribution and whether the mixed-effects model contains a neural network. Janssen et al. (2024) and Li et al. (2026) demonstrate the same point from opposite sides: Janssen et al. (2024) use a non-amortized estimation method, with no encoder neural network, to fit a model that does contain a neural network, whereas Li et al. (2026) use an amortized encoder neural network to fit a mechanistic system of ODEs that contains none. The variant presented in this paper is non-amortized and is applied both to a model that contains a neural network (Section 5.2) and to models that do not (Section 5.1, Appendix A.4), which covers the remaining corners. Whether a neural network appears in the model and whether one appears in the estimation method are therefore independent choices, and the base form of VEM fundamentally requires neither.

A similar separation holds between the VEM objective and how it is optimized. Although the objective is stochastic, the sample average approximation used here (Section 3.4) makes the per-iteration objective deterministic, so VEM does not require stochastic optimization algorithms to converge. This decouples the objective from the optimizer used to maximize it. The algorithm is also separate from how the gradient of the objective is computed, e.g. using forward-mode or reverse-mode AD (Appendix A.2). This choice does not change the definition of VEM, but it is very consequential for performance on large models. It is what allows VEM with reverse-mode AD to scale to a model on which forward-mode FOCE cannot complete a single iteration (Section 5.2), a distinction that has received little attention in related work.

## 5 Experiments

In this section, we present two experiments to evaluate the performance and scalability of the VEM algorithm compared to FOCE:

1. Warfarin model
    - A standard PK - PD model fitted to 31 subjects with a total of 464 observations.
    - VEM, FOCE and SAEM will be compared in terms of parameter estimates and log likelihoods.
2. DeepNLME Friberg model
    - An unnecessarily large NLME model incorporating neural networks, with 15,410 population parameters and 16 random effects in total.
    - A simulated dataset is used with 15 subjects and 28 observations per subject.

- VEM's scalability is evaluated over 500 iterations of VEM and compared to FOCE, reporting both the per-iteration runtime and the improvement in the FOCE-evaluated log likelihood.

In addition, Appendix A.4 presents an IRT model (a discrete-response, ordinal, longitudinal model) as an appendix experiment, demonstrating that VEM extends beyond Gaussian responses. That appendix also reports a Monte Carlo simulation study (Appendix A.4.3) assessing the accuracy and precision with which VEM recovers the data-generating population parameters, quantified through the coverage of its confidence intervals across simulated datasets.

These experiments aim to highlight the correctness and performance of VEM in both small and large-scale NLME modeling scenarios. The warfarin experiment was run using Pumas 2.8.1 and Julia 1.11.8 on a JuliaHub (a Julia cloud computing platform) virtual machine with the following configuration:

- OS: Linux (x86_64-linux-gnu)
- Host CPU: Intel(R) Xeon(R) Platinum 8375C @ 2.90GHz
- vCPUs allocated to the VM: 16
- Julia threads: 16
- RAM: 128 GB

We only used the deterministic VEM variant in these experiments. Other variants may be presented and compared in future work.

The DeepNLME experiment (Section 5.2) was run using a development version of Pumas and DeepPumas with Julia 1.11.9 on a JuliaHub virtual machine. This run executed on a machine with the same host CPU as above but a larger allocation:

- OS: Linux (x86_64-linux-gnu)
- Host CPU: Intel(R) Xeon(R) Platinum 8375C @ 2.90GHz
- vCPUs allocated to the VM: 64
- Julia threads: 16
- RAM: 512 GB

The larger virtual machine was selected for its memory rather than its core count, as reverse-mode automatic differentiation on the DeepNLME model is memory-intensive. Although 64 vCPUs were available, we allocated only 16 Julia threads to keep memory consumption low; 16 threads were sufficient given the size of the dataset.

The IRT experiment (Appendix A.4) was run using Pumas 2.8.1 and Julia 1.11.8 on the same JuliaHub virtual machine (the configuration listed above) as the warfarin experiment, so all reported runtimes are directly comparable. The SAEM comparison in the warfarin experiment (Table 1) was run in this same environment; SAEM is only available for models defined with Pumas's expectation–maximization model macro.

### 5.1 Warfarin model

The warfarin model (O'Reilly et al., 1963; O'Reilly & Aggeler, 1968) is a standard PK - PD model used to describe the dynamics of the warfarin drug in the body. The PK component of the model describes the drug concentration over time, while the PD component describes the drug's effect on a biological marker or response. The full model definition, including covariates, population parameters, individual parameters, random effects, dynamics, error model, and Pumas implementation, is provided in Appendix A.3.

### 5.1.1 Data

The Population Approach Group of Australia & New Zealand (PAGANZ) version of the warfarin dataset (O'Reilly & Aggeler, 1968) was used for this test. The dataset contains 31 subjects with a total of 464 observations for all subjects.

### 5.1.2 Comparison

In this section, we test the Pumas implementations of VEM and FOCE against each other, both using forward-mode AD on the warfarin model and dataset. For VEM, 15 samples per subject were used with presampling and a deterministic optimization algorithm (L-BFGS). A Gaussian variational family was used to approximate the posterior of the transformed and standardized random effects. A gradient norm tolerance of $10^{-3}$ was used for both VEM and FOCE. An additional relative objective change tolerance of $10^{-5}$ was used for both VEM and FOCE. The FOCE algorithm used a Newton trust region algorithm (Nocedal & Wright, 2006) for the inner optimization of the EBEs and a gradient norm tolerance of $10^{-5}$, which are the default settings. Note that the default values of the options may change between this publication and future versions of Pumas, so it is recommended to always check the documentation for the latest default values if the results in this section are to be reproduced. We additionally fit the same model with SAEM, started from the same typical parameter values as VEM and FOCE. SAEM was run with its default schedule of 200 exploratory, 100 middle, and 100 smoothing iterations. Following the recommendation to flatten the likelihood during the exploratory phase, the initial variability parameters were deliberately set well above their suspected final values: every random-effect standard deviation was initialized to 4, and the residual-error dispersions were likewise inflated. A fixed random seed was used for reproducibility.

The goal of this exercise is to compare the final parameter estimates and log likelihoods produced by the two algorithms and evaluate the convergence of the VEM algorithm. Both VEM and FOCE were initialized from the same parameter values, which are different from the optimal values. After the two estimation procedures completed, the Laplace method was used to approximate the log likelihoods at the two sets of parameter estimates. In the Laplace approximation, the EBEs were initialized from the variational distributions' modes for VEM and from the final FOCE EBEs for FOCE. The SAEM fit was re-scored under the Laplace method in the same way. Runtime is also reported but it is not the focus of this test. The runtime excludes the time it takes to calculate the log likelihood at the end.

Table 1 shows that the parameter estimates of VEM and FOCE and their log likelihoods are quite close, within about half a log-likelihood unit of each other. Some discrepancy between the two sets of estimates is nonetheless expected, because VEM and FOCE optimize slightly different objectives: VEM maximizes the ELBO, a lower bound on the marginal likelihood, whereas FOCE maximizes the first-order conditional approximation of it, so their optima need not coincide exactly. After the re-run, the most noticeable remaining differences between VEM and FOCE are in a few of the variance and residual-error components, namely the volume and absorption-time inter-individual variances in $\Omega_{\mathrm{PK}}$ (which VEM estimates somewhat smaller than FOCE) and the additive residual error of the PCA (prothrombin complex activity) response ($\sigma_{\mathrm{fx}}$, somewhat larger under VEM); these are among the parameters least constrained by the data. In runtime, VEM was significantly faster than FOCE in this case, although for small models it is possible to find options where FOCE is faster. SAEM recovered population estimates in the same range as VEM and FOCE, with a log likelihood about two and a half units below the other two. The largest discrepancies are in the absorption-related parameters ($t_{\mathrm{abs}}$ and lag, together with the $t_{\mathrm{abs}}$ inter-individual variance in $\Omega_{\mathrm{PK}}$ and $\omega_{\mathrm{lag}}$), and, to a lesser extent, in the PD potency and turnover parameters ($\mathrm{pop}_{\mathrm{c50}}$ and $\mathrm{pop}_{\mathrm{tover}}$); these are among the parameters least constrained by this dataset. On this example, SAEM was also the fastest of the three methods (Table 1). We caution, however, against drawing general conclusions about relative runtime from a single small model, as the ranking depends on the model, the dataset, and the algorithmic settings; a more extensive runtime benchmarking of SAEM against VEM and FOCE is out of scope for this paper and is left to future work.

Table 1: Parameter estimates, log likelihoods, and runtime comparison between VEM, FOCE, and SAEM using the warfarin model.

| | **VEM** | **FOCE** | **SAEM** |
|---|---|---|---|
| Log likelihood | −994.5 | −994.0 | −996.4 |
| Runtime (sec) | 49.6 | 124.2 | 22.0 |
| $\mathrm{pop}_{\mathrm{CL}}$ | 0.131 | 0.131 | 0.130 |
| $\mathrm{pop}_{\mathrm{V}}$ | 8.028 | 8.061 | 8.049 |
| $\mathrm{pop}_{\mathrm{tabs}}$ | 0.395 | 0.408 | 0.430 |
| $\mathrm{pop}_{\mathrm{lag}}$ | 0.994 | 1.024 | 0.816 |
| $\mathrm{pop}_{\mathrm{e0}}$ | 96.713 | 96.687 | 96.238 |
| $\mathrm{pop}_{\mathrm{emax}}$ | −1.046 | −1.060 | −1.115 |
| $\mathrm{pop}_{\mathrm{c50}}$ | 1.410 | 1.485 | 1.779 |
| $\mathrm{pop}_{\mathrm{tover}}$ | 13.688 | 13.961 | 14.876 |
| $\Omega_{\mathrm{PK}}$ | $\begin{pmatrix} 0.049 & 0.0 & 0.0 \\ 0.0 & 0.019 & 0.0 \\ 0.0 & 0.0 & 0.058 \end{pmatrix}$ | $\begin{pmatrix} 0.053 & 0.0 & 0.0 \\ 0.0 & 0.024 & 0.0 \\ 0.0 & 0.0 & 0.077 \end{pmatrix}$ | $\begin{pmatrix} 0.051 & 0.0 & 0.0 \\ 0.0 & 0.021 & 0.0 \\ 0.0 & 0.0 & 0.173 \end{pmatrix}$ |
| $\omega_{\mathrm{lag}}$ | 0.370 | 0.353 | 0.560 |
| $\Omega_{\mathrm{PD}}$ | $\begin{pmatrix} 0.002 & 0 & 0 & 0 \\ 0 & < 10^{-5} & 0 & 0 \\ 0 & 0 & 0.151 & 0 \\ 0 & 0 & 0 & 0.009 \end{pmatrix}$ | $\begin{pmatrix} 0.003 & 0 & 0 & 0 \\ 0 & 1.2 \times 10^{-4} & 0 & 0 \\ 0 & 0 & 0.163 & 0 \\ 0 & 0 & 0 & 0.009 \end{pmatrix}$ | $\begin{pmatrix} 0.003 & 0 & 0 & 0 \\ 0 & 7.5 \times 10^{-4} & 0 & 0 \\ 0 & 0 & 0.144 & 0 \\ 0 & 0 & 0 & 0.013 \end{pmatrix}$ |
| $\sigma_{\mathrm{prop}}$ | 0.076 | 0.076 | 0.075 |
| $\sigma_{\mathrm{add}}$ | 0.337 | 0.333 | 0.334 |
| $\sigma_{\mathrm{fx}}$ | 3.836 | 3.576 | 3.491 |

### 5.2 DeepNLME Friberg model

#### 5.2.1 Model definition

The DeepNLME (Korsbo et al., 2023) Friberg model used in this experiment is an extension of the Friberg myelosuppression model (Friberg et al., 2002), replacing some of the linear terms in the original model with neural networks to allow complex nonlinear relationships. The idea of incorporating a neural network in the dynamics of a Friberg model has been previously validated on a real dataset using a different DeepNLME model (Zare et al., 2024). The test model used here is different from the one in Zare et al. (2024), and it was not validated on a real dataset. In fact, its only purpose in this paper is to demonstrate the scalability of VEM to large NLME models with neural networks, which are expected to have a large number of parameters and random effects. The model is unnecessarily large for the sake of this demonstration, and it is not intended to be used in practice.

##### 5.2.1.1 Population parameters

The population parameters are:

- PK parameters:
  - Clearance (CL): $\mathrm{TVCL} > 0$
  - Central volume (VC): $\mathrm{TVVC} > 0$
  - Inter-compartmental clearance (Q): $\mathrm{TVQ} > 0$
  - Peripheral volume (VP): $\mathrm{TVVP} > 0$
  - Absorption rate constant (KA): $\mathrm{TVKA} > 0$
- PD parameters:
  - Mean transit time (MTT): $\mathrm{TVMTT} > 0$
  - Baseline circulating cells (CIRC0): $\mathrm{TVCIRC0} > 0$
  - Maximum drug effect parameter ($\alpha$): $0 < \mathrm{TVALPHA} < 1$
  - Drug effect offset: $c > 0$
- Inter-individual variability:
  - PK - PD covariance matrix: $\Omega$, a positive definite covariance matrix of size $8 \times 8$.
- Residual variability:
  - PK additive error: $\sigma_{\mathrm{pk,a}} > 0$
  - PK proportional error: $\sigma_{\mathrm{pk,p}} > 0$
  - PD additive error: $\sigma_{\mathrm{pd,a}} > 0$
  - PD proportional error: $\sigma_{\mathrm{pd,p}} > 0$
- Neural network parameters:
  - Neural network 1 (NN1): A Multi-Layer Perceptron (MLP) with 5 input features and 3 hidden layers of 60 neurons each, with hyperbolic tangent activation functions, followed by a single output neuron with an identity activation function. No bias terms were used in any of the layers of NN1.
  - Neural network 2 (NN2): An MLP with 6 input features and 3 hidden layers of 60 neurons each, with hyperbolic tangent activation functions, followed by a single output neuron with a softplus activation function. All layers of NN2 use bias terms.

The total number of population parameters in this model is 15,410.

The mechanistic population parameters were initialized at plausible values consistent with the data-generating Friberg model: $\mathrm{TVCL} = 4$, $\mathrm{TVVC} = 70$, $\mathrm{TVQ} = 4$, $\mathrm{TVVP} = 50$, $\mathrm{TVKA} = 1$, $\mathrm{TVMTT} = 125$, $\mathrm{TVCIRC0} = 5$, the drug-effect scale $\mathrm{TVALPHA} = 0.1$, and the drug-effect offset $c = 3$. The inter-individual covariance was initialized to $\Omega = 0.09 \cdot I_8$ (a standard deviation of $0.3$ on the log scale), and the residual-error dispersions to $\sigma_{\mathrm{pk,a}} = \sigma_{\mathrm{pd,a}} = 10^{-6}$ (additive) and $\sigma_{\mathrm{pk,p}} = \sigma_{\mathrm{pd,p}} = 0.2$ (proportional). The neural-network weights were initialized with Glorot (Xavier) normal initialization and the bias terms (present only in NN2) to zero.

#### 5.2.1.2 Random effects

The random effects include:

- 8 PK - PD random effects: $\eta \sim N(0, \Omega)$, where $\Omega$ is a positive definite covariance matrix.
- 8 neural network random effects: $\eta_{\mathrm{NN1}} \sim N(0, I_4)$ and $\eta_{\mathrm{NN2}} \sim N(0, I_4)$.

The model has a total of 16 random effects.

#### 5.2.1.3 Individual parameters

The individual parameters are derived from the population parameters and random effects:

- $\mathrm{CL} = \mathrm{TVCL} \cdot \exp(\eta_1)$
- $\mathrm{VC} = \mathrm{TVVC} \cdot \exp(\eta_2)$
- $\mathrm{Q} = \mathrm{TVQ} \cdot \exp(\eta_3)$
- $\mathrm{VP} = \mathrm{TVVP} \cdot \exp(\eta_4)$
- $\mathrm{KA} = \mathrm{TVKA} \cdot \exp(\eta_5)$
- $\mathrm{MTT} = \mathrm{TVMTT} \cdot \exp(\eta_6)$
- $\mathrm{CIRC0} = \mathrm{TVCIRC0} \cdot \exp(\eta_7)$
- $\alpha = \mathrm{logistic}(\mathrm{logit}(\mathrm{TVALPHA}) + \eta_8)$
- $k_{\mathrm{cp}} = Q / \mathrm{VC}$
- $k_{\mathrm{pc}} = Q / \mathrm{VP}$
- $k_{\mathrm{tr}} = 4/\mathrm{MTT}$

#### 5.2.1.4 Dynamics

The system of differential equations governing the dynamics is:

- PK dynamics:

$$\frac{d\mathrm{Depot}}{dt} = -\mathrm{KA} \cdot \mathrm{Depot},$$

$$\frac{d\mathrm{Central}}{dt} = \mathrm{KA} \cdot \mathrm{Depot} - \left(\frac{\mathrm{CL}}{\mathrm{VC}} + k_{\mathrm{cp}}\right) \cdot \mathrm{Central} + k_{\mathrm{pc}} \cdot \mathrm{Peripheral},$$

$$\frac{d\mathrm{Peripheral}}{dt} = k_{\mathrm{cp}} \cdot \mathrm{Central} - k_{\mathrm{pc}} \cdot \mathrm{Peripheral},$$

- PD dynamics:

$$
\begin{aligned}
\frac{d\text{Prol}}{dt} &= k_{\text{tr}} \cdot \text{Prol} \cdot \left( \frac{1 - \alpha \cdot \text{logistic}\big(\text{NN1}\big(\frac{\text{conc}}{100}, \eta_{\text{NN1}}\big) - c\big)}{10^{-8} + \text{NN2}\big(\frac{\text{Circ}}{100\ \text{CIRC0}}, \frac{\text{CIRC0}}{100}, \eta_{\text{NN2}}\big)} - 1 \right), \\
\frac{d\text{Transit1}}{dt} &= k_{\text{tr}} \cdot \text{Prol} - k_{\text{tr}} \cdot \text{Transit1}, \\
\frac{d\text{Transit2}}{dt} &= k_{\text{tr}} \cdot \text{Transit1} - k_{\text{tr}} \cdot \text{Transit2}, \\
\frac{d\text{Transit3}}{dt} &= k_{\text{tr}} \cdot \text{Transit2} - k_{\text{tr}} \cdot \text{Transit3}, \\
\frac{d\text{Circ}}{dt} &= k_{\text{tr}} \cdot \text{Transit3} - k_{\text{tr}} \cdot \text{Circ},
\end{aligned}
$$

where $\text{conc} = \text{Central}/\text{VC}$.

#### 5.2.1.5 Observations

The model includes the following observation models:

- PK observations:

$$
\text{PK} \sim N\Big(\text{conc}, \sigma_{\text{pk,a}}^2 + \big(\sigma_{\text{pk,p}} \cdot \text{conc}\big)^2\Big)
$$

- PD observations:

$$
\text{PD} \sim N\Big(\text{Circ}, \sigma_{\text{pd,a}}^2 + \big(\sigma_{\text{pd,p}} \cdot \text{Circ}\big)^2\Big)
$$

### 5.2.2 Data

The data were simulated using a standard Friberg model. The dataset contains 15 subjects with 14 PK observations and 14 PD observations for each subject. Each subject was given 7 doses of 10,000 units with the first dose given at time $t = 0$ and the remaining doses given every 24 hours. The observations spanned a total of 408 hours.

### 5.2.3 Goal of the experiment

Note that the goal of this section is not to properly fit and validate the DeepNLME model. The model used here is unreasonable given the size of the data and the true number of degrees of freedom of inter-individual variability in the data-generating model. The goal is instead to provide an easy medium to scale up the number of population parameters and random effects for testing the performance and scalability of the estimation algorithms. Therefore, we ran the VEM algorithm for 500 iterations, both to measure the time per iteration and to check that the objective continues to improve at this scale, and we attempted to run a single iteration of FOCE. Each iteration performs a gradient calculation, which is the most expensive part of the algorithm, so the time per iteration is a good proxy for the time per gradient calculation. Because the model is intentionally oversized relative to the dataset, its estimates carry no scientific meaning. What this experiment measures is the cost and the optimization progress of VEM at this model size, not the quality of a fitted DeepNLME model; applying VEM to genuinely useful large models is left to future work.

### 5.2.4 Results

We ran the VEM algorithm with reverse-mode AD using 32 samples per subject and presampling. The variational family was a multivariate Gaussian with a diagonal covariance matrix, approximating the posterior of the transformed and standardized random effects. The optimization algorithm used was L-BFGS. VEM completed 500 iterations in 78.5 hours, or 9.4 minutes per iteration. Including the Julia just-in-time compilation time of the 0th iteration with finite gradient checks, the total fit time was 78.7 hours, so the one-time compilation overhead was 12.0 minutes, about 0.25% of the total. Over these 500 iterations, the FOCE-evaluated log likelihood improved from $-876.9$ at the initial parameters to $-799.2$. All three log likelihoods quoted in this section are computed the same way, so that they are on a common scale: the population parameters are held fixed at the values being scored, the EBEs of the random effects are obtained by the FOCE inner optimization, and the resulting FOCE objective is reported. No population-parameter optimization is performed in this scoring step. For the VEM fit, the inner optimization is initialized from the modes of the fitted variational distributions; for the initial parameters and for the data-generating model, it is initialized from the Pumas default. The data-generating model's value is obtained by scoring the standard Friberg model that generated the data at the parameter values used to simulate it. The FOCE-evaluated log likelihood of the data-generating model on the same dataset is $-816.5$, so the fitted model ends up about 17 log-likelihood units *above* the model that generated the data. This is clear over-fitting rather than a better description of the data-generating process, and it is exactly what one should expect here: the DeepNLME model has 15,410 population parameters and 16 random effects, while the dataset provides only 420 observations from 15 subjects, and the data-generating Friberg model has far fewer degrees of freedom of inter-individual variability than the fitted one. With that many parameters relative to that little data, the likelihood can be driven well past that of the data-generating model by fitting noise. For the purposes of this experiment, what the log-likelihood trajectory establishes is that VEM is genuinely optimizing the objective at this scale, not that the resulting model is a good one. In contrast, the FOCE estimation algorithm with forward-mode AD did not complete a single iteration after 24 hours and was terminated. Note that reverse-mode AD requires significantly more upfront compilation time compared to forward-mode AD. However, this added time is a one-time cost that gets amortized over multiple iterations, especially when the number of parameters (population + subject-specific) in the model is large.

## 6 Discussion

The above results demonstrate the correctness of the VEM algorithm, in that it produces parameter estimates and log likelihoods close to those obtained by FOCE on the warfarin model, and its scalability with reverse-mode AD, in that it completes 500 iterations in a practical amount of time on a large and complex DeepNLME model while FOCE with forward-mode AD fails to complete even a single iteration within 24 hours. Over those 500 iterations, VEM improves the FOCE-evaluated log likelihood on this model from $-876.9$ to $-799.2$, past the $-816.5$ of the data-generating model; this is over-fitting, and it is what one should expect for a model with 15,410 population parameters fitted to 420 observations. The computational experiments reported here are not comprehensive benchmarks of VEM against competing methods and their software implementations. We plan to conduct such benchmarks in future work, and we hope that other researchers will also objectively evaluate VEM against competing methods and their software implementations.

Note that in Section 5.2, the comparison between VEM and FOCE reflects both the algorithmic differences between the two algorithms and the differences between reverse-mode and forward-mode AD. When estimating the population parameters with FOCE, forward-mode AD was used for multiple reasons. Estimating the population parameters with FOCE requires second-order derivatives of the individual conditional log likelihood with respect to the random effects and population parameters. Pure reverse-mode AD is generally not well-suited to computing higher-order derivatives efficiently. It is therefore unclear how or if reverse-mode AD can benefit FOCE. Since the current implementation of the widely popular FOCE algorithm in Pumas uses forward-mode AD and since the VEM algorithm proposed in this paper was also implemented in Pumas, the Pumas implementation of FOCE was used as a baseline in Section 5.2. However, it is arguably a limitation of the comparison that the FOCE vs VEM timing comparison reported in Section 5.2 reflects both the algorithmic differences between FOCE and VEM and the differences between forward-mode and reverse-mode AD.

While showing promising results above, the VEM algorithm is naturally not without limitations. We discuss the main ones here.

1. VEM is sensitive to the number of samples per subject used. The choice of 15 and 32 samples per subject in the experiments was somewhat arbitrary, and it is possible that using more or fewer samples could lead to different performance characteristics. It is recommended to experiment with different numbers of samples, e.g. different multiples of the number of random effects in the model.
2. VEM is sensitive to the seed of the pseudo-random number generator used to sample $\xi_i$ from the base distribution $b(\xi_i)$. It is recommended to experiment with different seeds and check the stability of the results. This dependence on the seed is not specific to VEM; it is shared by the other stochastic estimation methods discussed in Section 2, since the MCMC sampling in MCEM and SAEM likewise depends on the seed. VEM is therefore no more sensitive to the seed than these established methods, and the same practice of checking the stability of the estimates across seeds applies.
3. VEM is sensitive to the choice of variational family. The use of a multivariate Gaussian with a diagonal covariance matrix by default is for computational efficiency, but it may not capture complex dependencies between random effects. It is recommended to experiment with different variational families, such as a full covariance Gaussian, if the default family does not yield good results.
4. Reverse-mode AD incurs significant upfront compilation costs as well as higher memory usage and runtime overhead compared to forward-mode AD, which may make it less efficient for small models. For large models with stiff ODEs, the memory requirements of reverse-mode AD may become prohibitive, and further algorithmic refinements may be needed to reduce memory usage. Therefore, the choice of AD mode should be guided by the size and complexity of the model, as well as the computational resources available.
5. The variational distribution's mode is not guaranteed to be close to the true posterior mode, especially in cases where the true posterior is multimodal or has heavy tails. This means that a separate step is needed to calculate the EBEs at the end of the optimization to make individual predictions. This is not a major limitation, but it is an additional step that is not needed in FOCE, where the EBEs are directly optimized during the estimation.

6. VEM does not calculate a log likelihood approximation during the estimation, and a separate step is needed to approximate it. This is not a major limitation, but it is an additional step that is not needed in FOCE, where the log likelihood approximation is calculated during the estimation.

The above limitations do not discount the potential of the VEM algorithm to be a powerful tool for fitting large and complex NLME models, such as DeepNLME models, large joint biomarker models, and large QSP models fitted as NLME models.

## 7 Conclusion

In this paper, we presented a variant of the VEM algorithm for fitting NLME models, as implemented in the Pumas statistical software. The VEM algorithm addresses a critical bottleneck in modern pharmacometric modeling: the computational intractability of fitting large, complex NLME models with hundreds or thousands of population parameters and random effects.

Our key contributions include:

1. A detailed exposition of the VEM algorithm tailored specifically to NLME models, including its connections to related methods such as MCEM, SAEM, the Laplace method, and VAEs.
2. A practical implementation leveraging both forward-mode and reverse-mode automatic differentiation to achieve scalability, with reverse-mode AD being particularly crucial for models with many parameters relative to outputs.
3. Empirical validation on two contrasting scenarios: a small, well-established warfarin PK - PD model and a large-scale DeepNLME Friberg model with over 15,000 population parameters.

Empirical results demonstrated that VEM achieves parameter estimates and log likelihoods comparable to FOCE for small models like the warfarin model, while offering significant scalability advantages for larger models, such as the DeepNLME Friberg model. Notably, the warfarin experiment showed that VEM not only produces competitive results but can be substantially faster than FOCE. However, a larger set of test models is needed to make any conclusive remarks about performance comparisons for small models. On the other hand, the DeepNLME Friberg experiment demonstrated that VEM can tackle models where traditional methods become computationally prohibitive, completing iterations in reasonable time while FOCE could not complete even a single iteration within 24 hours, and improving the FOCE-evaluated log likelihood over 500 iterations to beyond that of the data-generating model. This DeepNLME model is deliberately over-parameterized for the size of its dataset. Its final log likelihood therefore reflects over-fitting, not model quality. What the experiment establishes is narrower: VEM optimizes the objective of a model of this size at a practical per-iteration cost. Applying VEM to large models that are genuinely useful is left to future work. VEM's scalability makes it well-suited for fitting large QSP models as regular NLME models, as well as fitting big joint biomarker models with tens or even hundreds of biomarkers simultaneously. Both of these applications will be explored in future work.

## References

Arruda, J., Schälte, Y., Peiter, C., Teplytska, O., Jaehde, U., & Hasenauer, J. (2024). An Amortized Approach to Non-Linear Mixed-Effects Modeling Based on Neural Posterior Estimation. *Proceedings of the 41st International Conference on Machine Learning (ICML)*, *235*, 1865–1901. https://proceedings.mlr.press/v235/arruda24a.html

Baaz, M., Sjöberg, A., & Jirstrand, M. (2026). The empirical Bayes variational autoencoder—a neural ODE approach for population modeling in pharmacology. *CPT: Pharmacometrics & Systems Pharmacology*, *15*(7), e70280. https://doi.org/10.1002/psp4.70280

Bauer, R. J. (2019). NONMEM tutorial part II: estimation methods and advanced examples. *CPT: Pharmacometrics & Systems Pharmacology*, *8*(8), 538–556. https://doi.org/10.1002/psp4.12422

Baydin, A. G., Pearlmutter, B. A., Radul, A. A., & Siskind, J. M. (2018). Automatic Differentiation in Machine Learning: a Survey. *Journal of Machine Learning Research*, *18*(153), 1–43. https://doi.org/10.48550/arXiv.1502.05767

Beskos, A., Pillai, N., Roberts, G., Sanz-Serna, J.-M., & Stuart, A. (2013). Optimal tuning of the hybrid Monte Carlo algorithm. *Bernoulli*, *19*(5A), 1501–1534. https://doi.org/10.3150/12-BEJ414

Bezanson, J., Edelman, A., Karpinski, S., & Shah, V. B. (2017). Julia: A Fresh Approach to Numerical Computing. *SIAM Review*, *59*(1), 65–98. https://doi.org/10.1137/141000671

Blei, D. M., Kucukelbir, A., & McAuliffe, J. D. (2017). Variational Inference: A Review for Statisticians. *Journal of the American Statistical Association*, *112*(518), 859–877. https://doi.org/10.1080/01621459.2017.1285773

Bleistein, N., & Handelsman, R. A. (1986). *Asymptotic Expansions of Integrals*. Dover Publications.

Burden, R. L., Faires, J. D., & Burden, A. M. (2015). *Numerical Analysis* (10th ed.). Cengage Learning.

Burroni, J., Domke, J., & Sheldon, D. (2024). Sample Average Approximation for Black-Box Variational Inference. *Proceedings of the Fortieth Conference on Uncertainty in Artificial Intelligence (UAI)*, *244*, 471–498. https://proceedings.mlr.press/v244/burroni24a.html

Cao, Y., Li, S., Petzold, L., & Serban, R. (2003). Adjoint Sensitivity Analysis for Differential-Algebraic Equations: The Adjoint DAE System and Its Numerical Solution. *SIAM Journal on Scientific Computing*, *24*(3), 1076–1089. https://doi.org/10.1137/S1064827501380630

Chen, R. T. Q., Rubanova, Y., Bettencourt, J., & Duvenaud, D. K. (2018). Neural Ordinary Differential Equations. *Advances in Neural Information Processing Systems (Neurips)*, *31*. https://proceedings.neurips.cc/paper/2018/hash/69386f6bb1dfed68692a24c8686939b9-Abstract.html

Cremer, C., Li, X., & Duvenaud, D. (2018). Inference Suboptimality in Variational Autoencoders. *Proceedings of the 35th International Conference on Machine Learning (ICML)*, *80*, 1078–1086. https://doi.org/10.48550/arXiv.1801.03558

Delyon, B., Lavielle, M., & Moulines, E. (1999). Convergence of a Stochastic Approximation Version of the EM Algorithm. *Annals of Statistics*, *27*(1), 94–128. https://doi.org/10.1214/aos/1018031103

Dempster, A. P., Laird, N. M., & Rubin, D. B. (1977). Maximum Likelihood from Incomplete Data via the EM Algorithm. *Journal of the Royal Statistical Society. Series B (Methodological)*, *39*(1), 1–38. https://www.jstor.org/stable/2984875

Embretson, S. E., & Reise, S. P. (2000). *Item Response Theory for Psychologists*. Lawrence Erlbaum Associates.

Freitas, N. de, F. R. Højen-Sørensen, P. A. d., Jordan, M. I., & Russell, S. J. (2001). Variational MCMC. *Proceedings of the 17th Conference on Uncertainty in Artificial Intelligence (UAI)*, 120–127.

Friberg, L. E., Henningsson, A., Maas, H., Nguyen, L., & Karlsson, M. O. (2002). Model of Chemotherapy-Induced Myelosuppression with Parameter Consistency across Drugs. *Journal of Clinical Oncology*, *20*(24), 4713–4721. https://doi.org/10.1200/JCO.2002.02.140

Friston, K., Mattout, J., Trujillo-Barreto, N., Ashburner, J., & Penny, W. (2007). Variational Free Energy and the Laplace Approximation. *Neuroimage*, *34*(1), 220–234. https://doi.org/10.1016/j.neuroimage.2006.08.035

Giordano, R., Ingram, M., & Broderick, T. (2024). Black Box Variational Inference with a Deterministic Objective: Faster, More Accurate, and Even More Black Box. *Journal of Machine Learning Research*, *25*(18), 1–39. https://doi.org/10.48550/arXiv.2304.05527

Glynn, P. W. (1990). Likelihood Ratio Gradient Estimation for Stochastic Systems. *Communications of the ACM*, *33*(10), 75–84. https://doi.org/10.1145/84537.84552

Griewank, A., & Walther, A. (2008). *Evaluating Derivatives: Principles and Techniques of Algorithmic Differentiation* (2nd ed.). SIAM. https://doi.org/10.1137/1.9780898717761

Hastings, W. K. (1970). Monte Carlo sampling methods using Markov chains and their applications. *Biometrika*, *57*(1), 97–109. https://doi.org/10.1093/biomet/57.1.97

Hindmarsh, A. C., Brown, P. N., Grant, K. E., Lee, S. L., Serban, R., Shumaker, D. E., & Woodward, C. S. (2005). SUNDIALS: Suite of Nonlinear and Differential/Algebraic Equation Solvers. *ACM Transactions on Mathematical Software*, *31*(3), 363–396. https://doi.org/10.1145/1089014.1089020

Hinton, G. E., & Salakhutdinov, R. R. (2006). Reducing the Dimensionality of Data with Neural Networks. *Science*, *313*(5786), 504–507. https://doi.org/10.1126/science.1127647

Hoffman, M. D., & Gelman, A. (2014). The No-U-Turn Sampler: Adaptively Setting Path Lengths in Hamiltonian Monte Carlo. *Journal of Machine Learning Research*, *15*(47), 1593–1623. http://jmlr.org/papers/v15/hoffman14a.html

Hotelling, H. (1933). Analysis of a complex of statistical variables into principal components. *Journal of Educational Psychology*, *24*(6), 417–441. https://doi.org/10.1037/h0071325

Janssen, A., Bennis, F. C., Cnossen, M. H., & Mathôt, R. A. A. (2024). Mixed Effect Estimation in Deep Compartment Models: Variational Methods Outperform First-Order Approximations. *Journal of Pharmacokinetics and Pharmacodynamics*, *51*(6), 797–808. https://doi.org/10.1007/s10928-024-09931-w

Jordan, M. I., Ghahramani, Z., Jaakkola, T. S., & Saul, L. K. (1999). An Introduction to Variational Methods for Graphical Models. *Machine Learning*, *37*(2), 183–233. https://doi.org/10.1023/A:1007665907178

Kahn, H., & Harris, T. E. (1951). Estimation of Particle Transmission by Random Sampling. *National Bureau of Standards Applied Mathematics Series*, *12*, 27–30.

Karimi, B., Lavielle, M., & Moulines, E. (2020). f-SAEM: A fast stochastic approximation of the EM algorithm for nonlinear mixed effects models. *Computational Statistics & Data Analysis*, *141*, 123–138. https://doi.org/10.1016/j.csda.2019.07.001

Kass, R. E., & Raftery, A. E. (1995). Bayes Factors. *Journal of the American Statistical Association*, *90*(430), 773–795. https://doi.org/10.1080/01621459.1995.10476572

Kingma, D. P., & Ba, J. (2015). Adam: A Method for Stochastic Optimization. *Proceedings of the 3rd International Conference on Learning Representations (ICLR)*. https://doi.org/10.48550/arXiv.1412.6980

Kingma, D. P., & Welling, M. (2013). Auto-Encoding Variational Bayes. *Arxiv Preprint Arxiv:1312.6114*. https://doi.org/10.48550/arXiv.1312.6114

Korsbo, N., Tarek, M., Elrod, C., Soubret, A., Brizzi, F., Rackauckas, C., Gobburu, J., & Ivaturi, V. (2023,). DeepPumas for automatic discovery of individualizable functions governing longitudinal patient outcomes. *Population Approach Group in Europe (PAGE)*.

Kuhn, E., & Lavielle, M. (2004). Coupling a stochastic approximation version of EM with an MCMC procedure. *ESAIM: Probability and Statistics*, *8*, 115–131. https://doi.org/10.1051/ps:2004007

Kuhn, E., & Lavielle, M. (2005). Maximum likelihood estimation in nonlinear mixed effects models. *Computational Statistics & Data Analysis*, *49*(4), 1020–1038. https://doi.org/10.1016/j.csda.2004.07.002

Lange, K., Hunter, D. R., & Yang, I. (2000). Optimization Transfer Using Surrogate Objective Functions. *Journal of Computational and Graphical Statistics*, *9*(1), 1–20. https://doi.org/10.2307/1390605

Li, Z., Prague, M., Thiébaut, R., & Clairon, Q. (2026). Variational autoencoder for inference of nonlinear mixed effect models based on ordinary differential equations. *Arxiv Preprint*. https://doi.org/10.48550/arXiv.2601.17400

Lindstrom, M. J., & Bates, D. M. (1990). Nonlinear Mixed Effects Models for Repeated Measures Data. *Biometrics*, *46*(3), 673–687. https://doi.org/10.2307/2532087

Louis, T. A. (1982). Finding the Observed Information Matrix When Using the EM Algorithm. *Journal of the Royal Statistical Society: Series B (Methodological)*, *44*(2), 226–233. https://doi.org/10.1111/j.2517-6161.1982.tb01203.x

Ma, Y., Dixit, V., Innes, M. J., Guo, X., & Rackauckas, C. (2021). A Comparison of Automatic Differentiation and Continuous Sensitivity Analysis for Derivatives of Differential Equation Solutions. *2021 IEEE High Performance Extreme Computing Conference (HPEC)*, 1–9. https://doi.org/10.1109/HPEC49654.2021.9622796

Metropolis, N., Rosenbluth, A. W., Rosenbluth, M. N., Teller, A. H., & Teller, E. (1953). Equation of State Calculations by Fast Computing Machines. *The Journal of Chemical Physics*, *21*(6), 1087–1092. https://doi.org/10.1063/1.1699114

Mogensen, P. K., Noack, A., Tarek, M., Elrod, C., Müller-Widmann, D., Krumbiegel, J., Hatherly, M., Korsbo, N., Rackauckas, C., & Ivaturi, V. (2026). Pumas: A Unified Platform for Pharmaceutical Modeling and Simulation. *QM*, *1*(2). https://doi.org/10.70534/qm-2026-0011

Neal, R. M. (2011). MCMC using Hamiltonian dynamics. In S. Brooks, A. Gelman, G. L. Jones, & X.-L. Meng (Eds.), *Handbook of Markov Chain Monte Carlo: Handbook of Markov Chain Monte Carlo* (pp. 113–162). Chapman, Hall/CRC. https://doi.org/10.1201/b10905

Nocedal, J., & Wright, S. J. (2006). *Numerical Optimization* (2nd ed.). Springer. https://doi.org/10.1007/978-0-387-40065-5

O'Reilly, R. A., & Aggeler, P. M. (1968). Studies on Coumarin Anticoagulant Drugs: Initiation of Warfarin Therapy Without a Loading Dose. *Circulation*, *38*(1), 169–177. https://doi.org/10.1161/01.CIR.38.1.169

O'Reilly, R. A., Aggeler, P. M., & Leong, L. S. (1963). Studies on the Coumarin Anticoagulant Drugs: The Pharmacodynamics of Warfarin in Man. *The Journal of Clinical Investigation*, *42*(10), 1542–1551. https://doi.org/10.1172/JCI104839

Pearson, K. (1901). On lines and planes of closest fit to systems of points in space. *The London, Edinburgh, And Dublin Philosophical Magazine and Journal of Science*, *2*(11), 559–572. https://doi.org/10.1080/14786440109462720

Ranganath, R., Gerrish, S., & Blei, D. M. (2014). Black Box Variational Inference. *Proceedings of the 17th International Conference on Artificial Intelligence and Statistics (AISTATS)*, *33*, 814–822. https://doi.org/10.48550/arXiv.1401.0118

Rezende, D. J., Mohamed, S., & Wierstra, D. (2014). Stochastic Backpropagation and Approximate Inference in Deep Generative Models. *Proceedings of the 31st International Conference on Machine Learning (ICML)*, 1278–1286. https://doi.org/10.48550/arXiv.1401.4082

Robert, C. P., & Casella, G. (2004). *Monte Carlo Statistical Methods* (2nd ed.). Springer. https://doi.org/10.1007/978-1-4757-4145-2

Roberts, G. O., Gelman, A., & Gilks, W. R. (1997). Weak convergence and optimal scaling of random walk Metropolis algorithms. *The Annals of Applied Probability*, *7*(1), 110–120. https://doi.org/10.1214/aoap/1034625254

Roeder, G., Wu, Y., & Duvenaud, D. (2017,). Sticking the Landing: Simple, Lower-Variance Gradient Estimators for Variational Inference. *Advances in Neural Information Processing Systems 30 (Neurips 2017)*.

Rohleff, J., Bachmann, F., Nahum, U., Bräm, D., Steffens, B., Pfister, M., Koch, G., & Schropp, J. (2025). Redefining Parameter Estimation and Covariate Selection via Variational Autoencoders: One Run Is All You Need. *CPT: Pharmacometrics & Systems Pharmacology*, *14*(12), 2232–2243. https://doi.org/10.1002/psp4.70129

Samejima, F. (1969). Estimation of Latent Ability Using a Response Pattern of Graded Scores. *Psychometrika Monograph Supplement*, *34*(4, Pt. 2), 1–97. https://doi.org/10.1007/BF03372160

Sheiner, L. B., & Beal, S. L. (1980). Evaluation of Methods for Estimating Population Pharmacokinetic Parameters. I. Michaelis-Menten Model: Routine Clinical Pharmacokinetic Data. *Journal of Pharmacokinetics and Biopharmaceutics*, *8*(6), 553–571. https://doi.org/10.1007/BF01060053

Tipping, M. E., & Bishop, C. M. (1999). Probabilistic Principal Component Analysis. *Journal of the Royal Statistical Society: Series B (Statistical Methodology)*, *61*(3), 611–622. https://doi.org/10.1111/1467-9868.00196

Wei, G. C. G., & Tanner, M. A. (1990). A Monte Carlo Implementation of the EM Algorithm and the Poor Man's Data Augmentation Algorithms. *Journal of the American Statistical Association*, *85*(411), 699–704. https://doi.org/10.2307/2290005

Williams, R. J. (1992). Simple Statistical Gradient-Following Algorithms for Connectionist Reinforcement Learning. *Machine Learning*, *8*(3–4), 229–256. https://doi.org/10.1007/BF00992696

Zakkour, A., Perret, C., & Slaoui, Y. (2023). Stochastic Expectation Maximization Algorithm for Linear Mixed-Effects Model with Interactions in the Presence of Incomplete Data. *Entropy*, *25*(3), 473. https://doi.org/10.3390/e25030473

Zare, R. N., Busse, D., Neldemo, I., Pitarch, A. P., Friberg, L. E., & Farnoud, A. (2024,). A Neural Networks-assisted NLME Framework: Case Study on Modeling Platelet Counts. *Population Approach Group in Europe (PAGE)*. https://doi.org/10.13140/RG.2.2.17690.86720

# A Appendix

## A.1 ELBO gradient details

Both the L-BFGS and Adam optimization algorithms require the gradient of the ELBO with respect to both $\theta$ and $\kappa$. The gradient of the population ELBO is the sum of the gradients of the individual ELBOs. So we focus on the individual ELBO:

$$\mathrm{ELBO}_i \approx \frac{1}{M} \sum_{j=1}^{M} G\big(T(\xi_{i,j}; \kappa_i); \theta, \kappa_i\big)$$

$$G(\eta_i; \theta, \kappa_i) = \log p(y_i \mid \eta_i, \theta) + \log p(\eta_i \mid \theta) - \log q(\eta_i; \kappa_i)$$

We can break each term in the summation into two sub-terms:

$$\text{ELBO}_i \approx \frac{1}{M} \sum_{j=1}^{M} (J_{i,j} - Q_{i,j})$$

where

$$\begin{aligned} J_{i,j} &= \log p\big(y_i \mid \eta_i = T(\xi_{i,j}; \kappa_i), \theta\big) + \log p\big(\eta_i = T(\xi_{i,j}; \kappa_i) \mid \theta\big) \\ Q_{i,j} &= \log q\big(\eta_i = T(\xi_{i,j}; \kappa_i); \kappa_i\big) \end{aligned} \tag{7}$$

$Q_{i,j}$ is only a function of $\kappa_i$, not $\theta$, and it often has an analytic gradient for simple variational families. For more complicated variational families, AD (Baydin et al., 2018) can be used to calculate the gradient of $Q_{i,j}$ with respect to $\kappa_i$.

$J_{i,j}$ is a function of both $\theta$ and $\kappa_i$. The partial derivative of $J_{i,j}$ with respect to $\theta$ can be easily calculated using AD. When computing the partial derivative of $J_{i,j}$ with respect to $\kappa_i$, the following chain rule is used:

$$\frac{dJ_{i,j}}{d\kappa_i} = \frac{d\eta_{i,j}}{d\kappa_i}^T \cdot \frac{\partial J_{i,j}}{\partial \eta_{i,j}} \tag{8}$$

where $\eta_{i,j} = T(\xi_{i,j}; \kappa_i)$.

$d\eta_{i,j}/d\kappa_i$ has a closed form for simple transformations $\eta_{i,j} = T(\xi_{i,j}; \kappa_i)$ and is generally tractable using AD for arbitrary invertible transformations. $\partial J_{i,j}/\partial \eta_{i,j}$ can be calculated using AD.

### A.2 Automatic differentiation

The VEM algorithm requires calculating multiple gradients as shown in Appendix A.1. To calculate these derivatives for an arbitrary NLME model, either numerical differentiation (Burden et al., 2015) or automatic differentiation (Baydin et al., 2018) can be used. We review the most common automatic and numerical differentiation methods in this section because the choice of differentiation algorithm greatly impacts the performance of the NLME model fitting algorithms.

#### A.2.1 Automatic differentiation vs numerical differentiation

AD, also known as *algorithmic differentiation* (Baydin et al., 2018; Griewank & Walther, 2008), is a family of techniques for computing derivatives of functions implemented as computer programs, provided the function is mathematically differentiable at the points of interest. The two practical variants are:

- Forward-mode AD, and
- Reverse-mode AD.

The main alternative families of differentiation techniques for computer programs are *numerical differentiation* (e.g., the finite difference method and the complex step method) and *symbolic differentiation*. Unlike numerical differentiation,

AD does not introduce truncation error: the only inaccuracy comes from the usual floating-point roundoff incurred by the underlying arithmetic. Unlike symbolic differentiation, AD does not construct or simplify a closed-form expression for the derivative; it propagates numerical derivative values alongside the function evaluation.

Finite difference methods treat the function as a black box, evaluating it at slightly perturbed real-valued inputs to estimate derivatives. The complex step method is similar, but perturbs the input in the complex plane, which reduces numerical errors but requires the function to accept complex inputs.

Symbolic differentiation constructs and manipulates mathematical expressions for the function and its derivatives, which can allow for simplification, but is often invasive and requires building an expression tree of all sub-functions and intermediate values.

Forward-mode and reverse-mode AD are the most widely used in practice for differentiating programs. They do not generally expand or simplify expressions, but instead propagate derivatives through the computational graph of a function. The key difference between forward-mode and reverse-mode is the direction in which derivatives are propagated.

### A.2.2 Manual derivative propagation

Forward-mode and reverse-mode AD automate the same derivative propagation one would carry out by hand via the chain rule, so we first describe that manual procedure before returning to AD.

Consider the simple case of a composition of three functions $G$, $F$, and $H$, each with one input vector and one output vector.

$$v \leftarrow H \leftarrow y \leftarrow F \leftarrow x \leftarrow G \leftarrow u$$

where $u$ is the input to the function composition and $v$ is its output. By the chain rule, the Jacobian $dv/du$ is:

$$\frac{dv}{du} = \frac{dv}{dy} \cdot \frac{dy}{dx} \cdot \frac{dx}{du}$$

To manually calculate the above Jacobian, derivatives can be either propagated from input to output, $v \leftarrow u$.

$$\underbrace{\frac{dv}{dy} \cdot \underbrace{\frac{dy}{dx} \cdot \frac{dx}{du}}_{dy/du}}_{dv/du}$$

Or they can be propagated from output to input, $v \rightarrow u$.

$$\underbrace{\underbrace{\frac{dv}{dy} \cdot \frac{dy}{dx}}_{dv/dx} \cdot \frac{dx}{du}}_{dv/du}$$

The former is called *forward* derivative propagation, and the latter is *reverse* derivative propagation.

Suppose $u$, $x$, and $y$ have size $n = 100$, and $v$ is a scalar.

$$\underbrace{\frac{dv}{du}}_{1\times 100} = \underbrace{\frac{dv}{dy}}_{1\times 100} \cdot \underbrace{\frac{dy}{dx}}_{100\times 100} \cdot \underbrace{\frac{dx}{du}}_{100\times 100}$$

In this case, reverse derivative propagation ($v \to u$) is more efficient. This is because at every step, we would be left multiplying a Jacobian matrix of one of the intermediate functions by a row vector of size $n = 100$. This matrix-vector multiplication has $O(n^2)$ time complexity and $O(n)$ memory complexity. Doing the multiplication in the forward direction ($v \leftarrow u$) requires more expensive matrix-matrix multiplications with $O(n^3)$ time complexity and $O(n^2)$ memory complexity. On the other hand, if the input $u$ is a scalar and the output $v$ has size $n = 100$, forward derivative propagation will be more efficient for the same reason, i.e., only matrix-vector multiplications are required.

In practice, the dimensions of intermediate variables vary along the chain, so the cost of either propagation order depends on the full sequence of intermediate sizes, not just the input and output dimensions. However, when the overall function has a small number of inputs and a large number of outputs, forward derivative propagation tends to be more efficient; conversely, when the overall function has a large number of inputs and a small number of outputs, reverse derivative propagation tends to be more efficient. Critically, this latter case is the regime of gradient computation: for a scalar-valued function, reverse propagation computes the entire gradient at a cost that is bounded by a small constant multiple (typically 3-5x) of the cost of evaluating the function itself, regardless of the input dimension. This is the well-known *cheap gradient principle* (Griewank & Walther, 2008).

For function compositions and functions with more complicated computational graphs than a linear function composition, the most efficient way to calculate the derivatives can involve a hybrid forward and reverse derivative propagation.

### A.2.3 Forward-mode vs reverse-mode AD

AD systems automatically build the chain of functions from the computer program and propagate derivatives up or down the chain:

- Forward-mode AD systems use forward derivative propagation.
- Reverse-mode AD systems use reverse derivative propagation. This is also called back-propagation in the machine learning literature.

Forward-mode AD is generally easier to implement because the propagation of derivatives happens in the same natural function execution order. Once we compute $y$ and $dy/du$, we can forget $x$ and $dx/du$. And once we compute $v$ and $dv/du$, we can forget $y$ and $dy/du$.

Reverse-mode AD is more complicated to implement because the propagation of derivatives happens in the *opposite direction* to the natural function execution order. We must therefore calculate and *save* all the intermediate values $x$, $y$, and $v$, in the so-called *forward pass* before we can start propagating derivatives in the *reverse pass*. The additional book-keeping makes reverse-mode AD more memory intensive, especially for long function chains with many intermediate variables, because all intermediate values need to be stored before beginning to propagate derivatives backward.

The two modes therefore have very different cost profiles. For a function with $n$ inputs and a scalar output, forward-mode requires $O(n)$ function evaluations to compute the gradient (one per input direction), while reverse-mode computes the gradient at a cost that is independent of $n$, but with higher per-operation overhead and additional memory proportional to the number of intermediate values. As a consequence, even when the input dimension is larger than the output dimension, forward-mode AD can still be faster than reverse-mode AD when both dimensions are small, because the reverse-mode overhead dominates. When the number of inputs is much larger than the number of outputs, however, the cheap gradient principle takes over and reverse-mode is asymptotically faster. This is crucial for scaling fitting algorithms of NLME models to hundreds of thousands or even millions of parameters and subjects.

### A.2.4 One variable at a time

Forward-mode AD systems propagate the derivatives with respect to one (or a few) inputs at a time; this count is called the chunk size. Reverse-mode AD systems propagate the derivatives of one (or a few) outputs at a time. In forward-mode AD, consider the derivative of the entire output $v$ with respect to a single scalar input variable $u_i$, which is a component of $u$.

$$\frac{dv}{du_i} = \underbrace{\frac{dv}{dy} \cdot \underbrace{\frac{dy}{dx} \cdot \frac{dx}{du_i}}_{dy/du_i}}_{dv/du_i}$$

Similarly, in reverse-mode AD, consider the derivative of a single output variable $v_i$ with respect to the entire input $u$.

$$\frac{dv_i}{du} = \underbrace{\underbrace{\frac{dv_i}{dy} \cdot \frac{dy}{dx}}_{dv_i/dx} \cdot \frac{dx}{du}}_{dv_i/du}$$

The forward (reverse) propagation of the derivatives above involves nothing more than right (left) multiplying the Jacobian of each function in the chain by some column (row) vector. Therefore the two fundamental operators needed for each intermediate function in the chain are:

1. Jacobian-Vector Product (JVP), also known as the *pushforward* operator: $t \to J \cdot t$ where $t$ (known as the *tangent*) is a column vector of the appropriate size. $t$ will have the same shape as the function's input. $J \cdot t$ will have the same shape as the function's output. This operator is used by forward-mode AD systems to propagate derivatives forward, without constructing the full intermediate Jacobians. Each primitive function in the AD system, e.g., algebraic functions, will have a JVP operator defined.
2. Vector-Jacobian Product (VJP), also known as the *pullback* (or *adjoint*) operator: $c \to c \cdot J$ where $c$ is a row vector of the appropriate size. Or equivalently, we can write the operator as $\tilde{c} \to J^T \cdot \tilde{c}$, where $\tilde{c} = c^T$ (known as the *cotangent*) is a column vector instead of a row vector. $\tilde{c}$ will have the same shape as the function's output.

$J^T \cdot \tilde{c}$ will have the same shape as the function's input. This operator is used by reverse-mode AD systems to propagate derivatives in reverse, without constructing the full intermediate Jacobians. Each primitive function in the AD system, e.g., algebraic functions, will have a VJP operator defined.

By the same argument given earlier for manual propagation: for functions with few inputs and many outputs, forward-mode AD is recommended; for functions with many inputs and few outputs, reverse-mode AD is recommended. As a concrete illustration, the warfarin model in Section 5.1 has only a handful of population parameters and a small number of random effects per subject, so each individual ELBO evaluation has a small input dimension; forward-mode AD with a chunk size > 1 is therefore competitive with or faster than reverse-mode for that model, which is why Section 5.1 uses forward-mode AD. By contrast, the DeepNLME Friberg model in Section 5.2 has more than 15,000 population parameters in the embedded neural network, so reverse-mode AD is essential to make a single ELBO gradient evaluation tractable.

### A.2.5 ODEs and AD

When the NLME model includes an ODE, both forward-mode and reverse-mode AD can be used to calculate the derivatives of the solution of the ODE in one of two main ways:

1. **Discretize-then-differentiate**: discretize (numerically solve) the original ODE and then apply AD to the discretized solution, or
2. **Differentiate-then-discretize**: propagate the derivatives through the continuous (unknown) exact solution of the ODE, which involves constructing and then numerically solving (discretizing) additional ODEs to propagate the derivatives through the original ODEs.

In forward-mode AD, the former is known as *discrete (forward) sensitivity analysis* (or *discrete local sensitivity analysis*), and the latter is known as the *continuous (forward) sensitivity analysis*. In reverse-mode AD, the former is known as the *discrete adjoint method*, and the latter is known as the *continuous adjoint method* (Cao et al., 2003; Chen et al., 2018; Hindmarsh et al., 2005).

The discrete sensitivity and adjoint methods involve computing exact derivatives (up to floating-point roundoff) of the approximate (discretized) solution of the original ODE, using forward-mode and reverse-mode AD, respectively. In contrast, the continuous methods formulate additional ODEs to propagate the derivatives through the original continuous ODE. Hypothetically, if we could solve the original ODE as well as the continuous sensitivity or adjoint ODEs exactly, we would get exact derivatives of the exact continuous solution of the original ODE, not just derivatives of a numerical solution (discretization) of the original ODE. In particular, we can evaluate the derivatives at any point in time, not just at the discrete time points of the numerical solution of the original ODE. In practice, however, we need to discretize both the original ODE as well as the continuous sensitivity and adjoint ODEs to solve them numerically, which can generally lead to gradients that are inconsistent with the discretized solution of the original ODE, as described in more detail below.

In the forward-mode case, the continuous sensitivity method solves an augmented ODE that includes the original states and additional sensitivity states that represent the tangents (derivatives of the original states with respect to the NLME model parameters). This larger augmented ODE is solved forward in time instead of the original ODE. While not inherently unstable, the augmented system typically has a much larger dimension than the original ODE and can exhibit different stiffness properties than the original ODE, which can make it more expensive to solve accurately. Accurately solving the augmented ODE amounts to a pushforward operator for the exact continuous solution of the original ODE, which allows forward-mode AD to propagate derivatives through the ODE solution.

In the reverse-mode case, the continuous adjoint method first solves the original ODE forward in time and then solves an adjoint ODE backward in time, where the adjoint states represent the cotangents (derivatives of the log likelihood with respect to the original ODE states). Accurately solving the adjoint ODE amounts to a pullback operator for the exact continuous solution of the original ODE. However, the reverse-time integration of the adjoint ODE can be numerically unstable for certain classes of ODEs, particularly stiff or strongly dissipative systems (Cao et al., 2003; Chen et al., 2018). The discrete adjoint method, by contrast, does not have this stability issue because it only requires applying AD to the discretized solution of the original ODE, without needing to solve any additional ODEs.

Discrete sensitivity and adjoint methods are generally easier to implement than their continuous counterparts. However, special care must be taken to ensure that the discretization of the original ODE is differentiable, which can be non-trivial for some families of numerical solvers, e.g., adaptive time-stepping solvers. Additionally, the main disadvantage of the discrete adjoint method is that it requires storing all intermediate values of the discretized solution of the original ODE during the forward pass, which can be memory intensive for long time horizons or stiff ODEs that require many small steps. On the other hand, the main advantage of the discrete methods is that they guarantee that the gradients computed are exact (up to floating-point roundoff) for the discretized/numerical solution of the original ODE. Since the numerical solution of the original ODE is what is actually used in the NLME model to make predictions and define the log likelihood, it can be desirable to have gradients that are consistent with the discretized solution of the original ODE. In contrast, continuous sensitivity and adjoint methods can produce gradients that are inconsistent with the discretized solution of the original ODE, which can lead to inaccurate gradients that may mislead the optimization algorithm even when the ODE solution itself is reasonably accurate. Empirical comparisons of different ways of differentiating ODE solutions can be found in Ma et al. (2021).

A further consideration in pharmacometric NLME models is that the ODE is typically driven by *dose events*: discrete times at which the state vector jumps (e.g., a bolus dose) or the input rate changes (e.g., the start or end of an infusion). These event times may themselves depend on covariates or, in some cases, on parameters. Differentiating through events requires propagating sensitivities (or adjoint variables) across the discontinuities using event-handling jump conditions. Both the discrete and continuous methods can be extended to handle events, but the bookkeeping is non-trivial.

In this paper and in Pumas, we use the discrete sensitivity and adjoint methods for forward-mode and reverse-mode AD, respectively. This behavior may be changed in future versions of Pumas.

### A.3 Warfarin model definition

This section provides the full definition of the warfarin model (O'Reilly et al., 1963; O'Reilly & Aggeler, 1968) used in Section 5.1.

### A.3.1 Covariates

The model has the following covariates:

- WT: baseline weight in kg
- $\text{FSZV} = \text{WT} / 70$
- $\text{FSZCL} = (\text{WT} / 70)^{0.75}$

### A.3.2 Population parameters

The model has the following population parameters:

- Clearance (L/h/70kg): $\text{pop}_{\text{CL}} > 0.0$
- Central Volume (L/70kg): $\text{pop}_{\text{V}} > 0.0$
- Absorption time (h): $\text{pop}_{\text{tabs}} > 0.0$
- Lag time (h): $\text{pop}_{\text{lag}} > 0.0$
- Baseline: $\text{pop}_{\text{e0}} > 0.0$
- Emax: $\text{pop}_{\text{emax}}$
- EC50: $\text{pop}_{\text{c50}} > 0.0$
- Turnover: $\text{pop}_{\text{tover}} > 0.0$
- Inter-individual variability:
  - $\Omega_{\text{PK}}$: diagonal positive definite matrix of size $3 \times 3$
  - $\Omega_{\text{PD}}$: diagonal positive definite matrix of size $4 \times 4$
  - $\omega_{\text{lag}} > 0.0$
- Residual variability:
  - $\sigma_{\text{prop}} > 0.0$
  - $\sigma_{\text{add}} > 0.0$
  - $\sigma_{\text{fx}} > 0.0$

### A.3.3 Individual parameters

The model has the following individual parameters:

- PK parameters:
  - Clearance

$$\text{CL} = \text{FSZCL} \cdot \text{pop}_{\text{CL}} \cdot \exp(\eta_{\text{CL}})$$

  - Central volume

$$V_c = \text{FSZV} \cdot \text{pop}_{\text{V}} \cdot \exp(\eta_{\text{V}})$$

  - Absorption time

$$t_{\text{abs}} = \text{pop}_{\text{tabs}} \cdot \exp(\eta_{\text{tabs}})$$

- Absorption rate constant

$$K_a = \log(2)/t_{\text{abs}}$$

- Lag time

$$t_{\text{lag}} = \text{pop}_{\text{lag}} \cdot \exp(\eta_{\text{lag}})$$

- PD parameters:
  - Baseline: $e_0 = \text{pop}_{\text{e0}} \cdot \exp(\eta_{\text{e0}})$
  - Maximum effect: $E_{\max} = \text{pop}_{\text{emax}} \cdot \exp(\eta_{\text{emax}})$
  - Half-maximal concentration: $\text{EC}_{50} = \text{pop}_{\text{c50}} \cdot \exp(\eta_{\text{c50}})$
  - Turnover time: $t_{\text{over}} = \text{pop}_{\text{tover}} \cdot \exp(\eta_{\text{turn}})$
  - Elimination rate constant: $K_{\text{out}} = \log(2)/t_{\text{over}}$
  - Production rate: $R_{\text{in}} = e_0 \cdot K_{\text{out}}$

### A.3.4 Random effects

The model has the following random effects:

- PK random effects: $\eta_{\text{CL}}, \eta_{\text{V}}, \eta_{\text{tabs}} \sim N(0, \Omega_{\text{PK}})$
- Lag time random effect: $\eta_{\text{lag}} \sim N(0, \omega_{\text{lag}})$
- PD random effects: $\eta_{\text{e0}}, \eta_{\text{emax}}, \eta_{\text{c50}}, \eta_{\text{turn}} \sim N(0, \Omega_{\text{PD}})$

### A.3.5 Dynamics

The system of differential equations governing the dynamics is:

$$\frac{d\text{Depot}}{dt} = -K_a \cdot \text{Depot}$$

$$\frac{d\text{Central}}{dt} = K_a \cdot \text{Depot} - \text{CL} \cdot \frac{\text{Central}}{V_c}$$

$$\frac{d\text{Turnover}}{dt} = R_{\text{in}} \cdot \left(1 + E_{\max} \cdot \frac{\text{Central }/V_c}{\text{EC}_{50} + \text{Central }/V_c}\right) - K_{\text{out}} \cdot \text{Turnover}$$

where:

- $K_a$ is the absorption rate constant,
- CL is the clearance,
- $V_c$ is the central volume of distribution,
- $R_{\text{in}}$ is the zero-order production rate constant of the response,
- $K_{\text{out}}$ is the first-order elimination rate constant of the response,
- $E_{\max}$ is the maximum drug effect, and
- $\text{EC}_{50}$ is the drug concentration at which 50% of the maximum effect is achieved.

The dose was administered at time $t = 0$ to the depot compartment with a lag time of $t_{\text{lag}}$.

### A.3.6 Error model

The model uses the following error model:

- Warfarin concentration

$$\text{Conc} \sim N\left(\frac{\text{Central}}{V_c}, \left(\sigma_{\text{prop}} \cdot \frac{\text{Central}}{V_c}\right)^2 + \sigma_{\text{add}}^2\right)$$

- PCA (prothrombin complex activity; distinct from the principal component analysis abbreviation used in Section 3.10)

$$\text{PCA} \sim N\left(\text{Turnover}, \sigma_{\text{fx}}^2\right)$$

### A.3.7 Pumas implementation

The Pumas definition of the warfarin model used in this test is given below:

```
model = @model begin
    @param begin
        pop_CL ∈ RealDomain(lower = 0.0, init = 0.134)
        pop_V ∈ RealDomain(lower = 0.0, init = 8.11)
        pop_tabs ∈ RealDomain(lower = 0.0, init = 0.523)
        pop_lag ∈ RealDomain(lower = 0.0, init = 0.1)
        pop_e0 ∈ RealDomain(lower = 0.0, init = 100.0)
        pop_emax ∈ RealDomain(init = -1.0)
        pop_c50 ∈ RealDomain(lower = 0.0, init = 1.0)
        pop_tover ∈ RealDomain(lower = 0.0, init = 14.0)
        pk_Ω ∈ PDiagDomain([0.01, 0.01, 0.01])
        lag_ω ∈ RealDomain(lower = 0.0, init = 0.1)
        pd_Ω ∈ PDiagDomain([0.01, 0.01, 0.01, 0.01])
        σ_prop ∈ RealDomain(lower = 0.0, init = 0.00752)
        σ_add ∈ RealDomain(lower = 0.0, init = 0.0661)
        σ_fx ∈ RealDomain(lower = 0.0, init = 0.01)
    end
    @random begin
        pk_η ~ MvNormal(log.([pop_CL, pop_V, pop_tabs]), pk_Ω)
        lag_η ~ Normal(0.0, lag_ω)
        pd_η ~ MvNormal(log.([pop_e0, 1, pop_c50, pop_tover]), pd_Ω)
    end
```

```
    @covariates FSZV FSZCL
    @pre begin
        # PK
        CL = FSZCL * exp(pk_η[1])
        Vc = FSZV * exp(pk_η[2])
        tabs = exp(pk_η[3])
        Ka = log(2) / tabs
        # PD
        e0 = exp(pd_η[1])
        emax = pop_emax * exp(pd_η[2])
        c50 = exp(pd_η[3])
        tover = exp(pd_η[4])
        kout = log(2) / tover
        rin = e0 * kout
    end
    @dosecontrol begin
        lags = (Depot = pop_lag * exp(lag_η),)
    end
    @init begin
        Turnover = e0
    end
    @vars begin
        cp := Central / Vc
        ratein := Ka * Depot
        pd := 1 + emax * cp / (c50 + cp)
    end
    @dynamics begin
        Depot' = -ratein
        Central' = ratein - CL * cp
        Turnover' = rin * pd - kout * Turnover
    end
    @derived begin
        conc ~ @. Normal(cp, sqrt((σ_prop * cp)^2 + σ_add^2))
        pca ~ @. Normal(Turnover, σ_fx)
    end
end
```

### A.3.8 Expectation–maximization (SAEM) implementation

SAEM in Pumas requires a model defined with the expectation–maximization model macro (@emmodel) rather than the standard @model macro. The @emmodel below is the equivalent formulation of the warfarin model above, used for the SAEM fit in Section 5.1. It follows the same per-subject LogNormal parameterization as the @model: each typical value is the location of a LogNormal random effect, the covariate size factors FSZCL and FSZV are applied in @pre, and the negative multiplicative maximum effect is reproduced as $E_{\max} = -\mathrm{negemax}$, where negemax is itself a LogNormal random effect. The log likelihood reported for SAEM in Table 1 is obtained by re-scoring this fit under the Laplace approximation, placing it on the same scale as the VEM and FOCE fits.

```
wf_em = @emmodel begin
  @random begin
    CL      ~ 1 | LogNormal
    Vc      ~ 1 | LogNormal
    tabs    ~ 1 | LogNormal
    lag     ~ 1 | LogNormal
    e0      ~ 1 | LogNormal
    negemax ~ 1 | LogNormal
    c50     ~ 1 | LogNormal
    tover   ~ 1 | LogNormal
  end
  @covariates FSZV FSZCL
  @pre begin
    CLi  = FSZCL * CL
    Vci  = FSZV * Vc
    Ka   = log(2) / tabs
    Emax = -negemax
    Kout = log(2) / tover
    Rin  = e0 * Kout
  end
  @dosecontrol begin
    lags = (; Depot = lag)
  end
  @init begin
    Turnover = e0
  end
  @dynamics begin
    Depot'    = -Ka * Depot
```

```
        Central'  = Ka * Depot - CLi * (Central / Vci)
        Turnover' = Rin * (1 + Emax * (Central / Vci) / (c50 + Central / Vci)) - Kout * Turnover
    end
    @post begin
        cp = Central / (FSZV * Vc)
    end
    @error begin
        conc ~ CombinedNormal(cp)
        pca  ~ Normal(Turnover)
    end
end
```

The initial values used for the SAEM fit are given below. Typical values enter directly (the LogNormal base is the typical value); Ω is the tuple of per-random-effect standard deviations, and σ collects the residual-error dispersions: $(\text{additive}, \text{proportional})$ for the CombinedNormal concentration error and $(\text{additive}, )$ for the Normal PCA error. The variability initial values are deliberately large so that SAEM flattens the likelihood during its exploratory phase.

```
saem_init = (
    CL = 0.134, Vc = 8.11, tabs = 0.523, lag = 0.1,
    e0 = 100.0, negemax = 1.0, c50 = 1.0, tover = 14.0,
    Ω = (4.0, 4.0, 4.0, 4.0, 4.0, 4.0, 4.0, 4.0),
    σ = ((2.0, 0.5), (10.0,)),
)
```

### A.4 IRT model

This appendix presents an additional experiment that applies VEM to a discrete-response model, complementing the Gaussian-response experiments in Section 5 and showing that VEM is not restricted to Gaussian observations. We use an IRT model (Embretson & Reise, 2000), in which each subject answers a set of ordinal-scored items over time and a latent, time-varying ability governs the response probabilities. This is a longitudinal generalized (non-Gaussian) mixed model.

The model has two items, each scored on a four-category ordinal scale. A subject's latent ability is linear in time,

$$f_i(t) = f_{0,i} + s_i \cdot t, \quad f_{0,i} \sim N\left(\mu_f, \omega_f^2\right), \quad s_i \sim N\left(\mu_s, \omega_s^2\right),$$

with a per-subject random intercept $f_{0,i}$ and slope $s_i$. For item $j$ with ordered category cut-offs $b_j$ and discrimination $a_j$, the ordinal response uses a proportional-odds (ordered-logit) link, i.e. the graded response model (Samejima, 1969),

$$\text{score}_{ij}(t) \sim \text{OrderedLogit}\left(a_j \cdot \left(b_j - f_i(t)\right)\right).$$

The population parameters are the ability-trajectory means and standard deviations ($\mu_f, \mu_s, \omega_f, \omega_s$), the item cut-offs ($b_1, b_2$, parameterized through strictly increasing transforms), and the free discrimination ($a$; the first item's discrimination is fixed to 1 for identifiability). Each subject contributes two random effects, on the ability intercept and slope.

### A.4.1 Data

We simulate a dataset of 300 subjects, each answering the two items at five time points ($t = 0, 4, 8, 12, 24$). The data-generating parameter values are $\mu_f = -0.3, \omega_f = 0.8, \mu_s = 0.04, \omega_s = 0.02$, item-1 cut-offs $b_1 = (0, 1.5, 2.5)$, item-2 cut-offs $b_2 = (-1, 0.3, 1.2)$, and item-2 discrimination $a = 0.75$ (item-1 discrimination fixed at 1), and each subject's random intercept and slope are drawn from the distributions above.

### A.4.2 Comparison

We fit the model with VEM and, as an accuracy reference, with the exact Laplace approximation (LaplaceI). As in the warfarin experiment, the log likelihood of each fit is reported by re-evaluating it under the Laplace approximation at the fitted population parameters, placing the two methods on a common scale. VEM used the deterministic (pre-sampled) variant with a Gaussian variational family (diagonal covariance, the Pumas default), forward-mode AD, and the L-BFGS optimizer. Both fits were started from the same initial estimates, chosen away from the data-generating values: ability means $\mu_f = 0.5$, $\mu_s = 0.2$; ability standard deviations $\omega_f = 0.4$, $\omega_s = 0.1$; ordinal cut-offs initialized to modestly increasing values (both items starting from approximately $(0, 0.61, 1.21)$); and item-2 discrimination $a = 0.5$ (item-1 discrimination fixed at 1).

Table 2 shows that VEM and the exact Laplace approximation agree closely on this discrete-response model. The VEM log likelihood, evaluated under the Laplace approximation, is within about one unit of the LaplaceI optimum, and the population estimates (ability means, item cut-offs, and discrimination) match between the two methods and recover the data-generating values. The one exception is the slope standard deviation $\omega_s$, a small variance component (data-generating value 0.02): VEM shrinks it toward zero ($\approx 0.003$), whereas LaplaceI recovers $\approx 0.022$. Overall, the experiment demonstrates that VEM extends to discrete-response (ordinal) longitudinal models and recovers essentially the same fit as the exact Laplace approximation.

Table 2: Parameter estimates, log likelihoods (evaluated under the Laplace approximation at the fitted population parameters), and runtime for the IRT model, comparing VEM against the exact Laplace approximation (LaplaceI) on 300 subjects with two four-category ordinal items. The cut-off vectors $b_1, b_2$ are the strictly increasing ordered category thresholds; item $1'$s first threshold and discrimination are fixed for identifiability.

| | LaplaceI | VEM |
|---|---|---|
| Log likelihood | −3650.4 | −3651.2 |
| Runtime (sec) | 12.6 | 18.7 |
| $\mu_f$ | −0.374 | −0.379 |
| $\mu_s$ | 0.044 | 0.044 |
| $\omega_f$ | 0.730 | 0.764 |
| $\omega_s$ | 0.0221 | 0.0027 |
| $a$ (item 2 discrimination) | 0.915 | 0.911 |
| $b_1$ (item 1 cut-offs) | (0, 1.51, 2.42) | (0, 1.50, 2.40) |
| $b_2$ (item 2 cut-offs) | (-0.72, 0.28, 0.97) | (-0.72, 0.28, 0.96) |

### A.4.3 Simulation study

The comparison above uses a single dataset at fixed data-generating values. To assess how reliably VEM recovers the population parameters and how well-calibrated its confidence intervals are, we turn the experiment into a Monte Carlo simulation study. For each of $R = 500$ replicates we draw a fresh set of true population parameters from distributions centered on the values above, simulate an ordinal longitudinal dataset of $N = 200$ subjects at the same five time points, fit the model with VEM, form 95% Wald confidence intervals for the population parameters, and record for each parameter whether its interval contains the true value. The variance components are drawn away from zero so that they remain estimable ($\omega_f \sim \mathrm{Uniform}(0.5, 1.0)$, $\omega_s \sim \mathrm{Uniform}(0.05, 0.10)$); the ability means, item cut-offs, and discrimination are drawn around the values listed above. The reported coverage is the fraction of replicates whose interval covered the truth, which for a well-calibrated 95% interval should be close to 95%.

VEM does not provide an asymptotic covariance of its own, so we obtain standard errors from the exact-Laplace observed information evaluated at the VEM estimate (a zero-iteration LaplaceI fit started at the VEM coefficients). Because the VEM optimum is not a stationary point of the Laplace objective, the VEM estimates of $\omega_f$ and $\omega_s$ give an ill-conditioned Laplace-based variance-covariance estimate, with the ill-conditioning carried by these variance-component directions; we therefore hold $\omega_f$ and $\omega_s$ fixed at their VEM values when forming the covariance and report coverage for the remaining structural parameters (the ability means, item cut-offs, and discrimination). This is a pragmatic device. One remedy is to obtain the standard errors by bootstrapping instead, which avoids relying on the Laplace-based covariance at the cost of substantially more computation. Another is Louis' method (Louis, 1982), the covariance appropriate to an EM-type algorithm such as VEM, which recovers the observed information matrix from the complete-data score and information but is not currently implemented in Pumas. We leave to future work a comparison of these different confidence-interval estimation methods after a VEM fit.

Table 3 reports coverage over the 497 of 500 replicates for which the covariance could be formed (the remaining three yielded a non-positive-definite information). Coverage is close to nominal for the ability means (about 94%) and mildly below nominal for the item cut-off and discrimination parameters (about 88–90%), with small biases throughout. That the observed coverage falls below the nominal 95% level is most plausibly an artifact of how the intervals are constructed rather than a deficiency of VEM's estimates: the intervals are formed from a Laplace approximation to the observed information, evaluated at the VEM estimate after fitting with an EM-type method. This is consistent with the mild under-coverage of the cut-off and discrimination parameters, and we would expect an EM-appropriate covariance such as Louis' method (Louis, 1982) to bring coverage closer to nominal. A full investigation of this effect is left to future work.

Table 3: Coverage of 95% Wald confidence intervals and mean bias for the IRT population parameters over a simulation study of $R = 500$ replicates ($N = 200$ subjects each), fit with VEM. Coverage is the fraction of the 497 replicates with a usable covariance whose interval contained the data-generating value. The cut-off parameters $\theta_{b_1,\cdot}, \theta_{b_2,\cdot}$ are the strictly-increasing transform parameters on which the ordered cut-offs are estimated; the variance components $\omega_f, \omega_s$ are held fixed for the standard-error computation and are not assessed (see text). Bias is reported on the estimation scale.

| **Parameter** | **Coverage** | **Mean bias** |
|---|---|---|
| $\mu_f$ (ability-intercept mean) | 93.8% | $+0.011$ |
| $\mu_s$ (ability-slope mean) | 93.8% | $-0.001$ |
| $\theta_{b_1,1}$ (item-1 cut-off) | 88.7% | $-0.035$ |
| $\theta_{b_1,2}$ (item-1 cut-off) | 93.2% | $-0.015$ |
| $\theta_{b_2,1}$ (item-2 cut-off) | 89.9% | $+0.020$ |
| $\theta_{b_2,2}$ (item-2 cut-off) | 89.3% | $-0.033$ |
| $\theta_{b_2,3}$ (item-2 cut-off) | 87.5% | $-0.042$ |
| $a$ (item-2 discrimination) | 89.3% | $+0.016$ |
| Pooled | 90.7% | – |